\documentclass[manuscript]{aastex}

\usepackage{amsmath}
\usepackage{bm}
\usepackage{graphicx}

\shorttitle{1D Flare Model}
\shortauthors{Takasao et al.}

\begin{document}


\title{MAGNETOHYDRODYNAMIC SHOCKS IN AND ABOVE POST-FLARE LOOPS: TWO-DIMENSIONAL SIMULATION AND A SIMPLIFIED MODEL}


\author{Shinsuke Takasao\altaffilmark{1},
Takuma Matsumoto\altaffilmark{2},
Naoki Nakamura\altaffilmark{1}, and
Kazunari Shibata\altaffilmark{1}}

\email{takasao@kwasan.kyoto-u.ac.jp}
\altaffiltext{1}{Kwasan and Hida Observatories, Kyoto University, Yamashina, Kyoto 607-8471, Japan}
\altaffiltext{2}{Institute of Space and Astronautical Science (ISAS), Japan Aerospace Exploration Agency (JAXA), 3-1-1 Yoshinodai, Sagamihara, Kanagawa 252-5210, Japan}




\begin{abstract}
Solar flares are an explosive phenomenon, where super-sonic flows and shocks are expected in and above the post-flare loops. To understand the dynamics of post-flare loops, a two-dimensional magnetohydrodynamic (2D MHD) simulation of a solar flare has been carried out. We found new shock structures in and above the post-flare loops, which were not resolved in the previous work by \citet{yokoyama2001}. To study the dynamics of flows along the reconnected magnetic field, kinematics and energetics of the plasma are investigated along selected field lines. It is found that shocks are crucial to determine the thermal and flow structures in the post-flare loops. On the basis of the 2D MHD simulation, we have developed a new post-flare loop model which we call the pseudo-2D MHD model. The model is based on the 1D MHD equations, where all the variables depend on one space dimension and all the three components of the magnetic and velocity fields are considered. Our pseudo-2D model includes many features of the multi-dimensional MHD processes related to magnetic reconnection (particularly MHD shocks), which the previous 1D hydrodynamic models are not able to include. We compare the shock formation and energetics of a specific field line in the 2D calculation with those in our pseudo-2D MHD model, and we found that they give similar results. This model will allow us to study the evolution of the post-flare loops in a wide parameter space without expensive computational cost and without neglecting important physics associated with magnetic reconnection.
\end{abstract}


\section{Introduction}
Solar flares are a phenomenon in which a huge amount of magnetic energy stored in the coronal field is rapidly released by magnetic reconnection. Where magnetic reconnection is a physical process in which a magnetic field in a highly conducting plasma changes its connectivity due to finite resistivity. The rapid energy release is accompanied by various plasma dynamics, like shock waves and the chromospheric evaporation. For more details about solar flares, the reader is referred to \citet{shibata2011}.

Flares similar to solar flares have been observed from many stars and other astrophysical objects \citep[e.g.][]{koyama1994,tsuboi1998,gudel2004}. Since many of those flares and solar flares have many common features, solar flare models have been applied to those astrophysical flares \citep[e.g.][]{hayashi1996,shibata2002,machida2003,masada2010}. Note that there should be differences in the plasma parameters between those flares and solar flares, like the plasma density, Alfven speed, and system size. Since these values determine the time scales of different processes, this can make the evolution of different flares distinct. Therefore, to comprehensively understand the flare physics in the universe, we need to explore various flares in a wide range of the plasma parameters.

There are many attempts to model solar and stellar flare loops. One-dimensional (1D) hydrodynamic models have been widely developed to study the thermal evolution and flows in a single loop. Depending on the assumed main energy transfer process, the models can be categorized into two branches: conduction-heating models \citep[e.g.][]{nagai1980,peres1993} and electron-beam-heating models \citep[e.g.][]{fisher1985,mariska1989}. In general, the results of both models agree with observed emissions well. \citet{hori1997,hori1998} developed a pseudo-2D loop model based on 1D hydrodynamic calculations, and gave a simple description of observed soft X-ray emissions (see \citet{warren2006} for further development of this kind of models).

The first 2D magnetohydrodynamic (MHD) simulation of a solar flare based on a reconnection model with the heat conduction was performed by \citet{yokoyama2001}, and they studied the thermal evolution in the post-flare loops and the physics that determines the flare temperature. \citet{shiota2005} performed MHD simulations of reconnection with the heat conduction to study the coronal mass ejections and associated giant arcade, where they do not include the chromosphere. \citet{miyagoshi2004} carried out a MHD simulation of chromospheric evaporation jets formed as a result of reconnection between emerging magnetic flux and a coronal ambient field. Recently \citet{longcope2009} and \citet{longcope2014} developed a theoretical model in which the dynamics of a reconnected field line is considered. 

Shocks in and above the post-flare loops can be important for both the non-thermal particle acceleration and evolution of the thermal structure. The above-the-looptop hard X-ray source found by \citet{masuda1995} has posed problems about the electron acceleration mechanism (for recent progress, see \citet[e.g.][]{krucker2010} and \citet{oka2015}). One of the promising scenarios is the acceleration by fast shocks which are expected to be formed above the soft X-ray post-flare loops \citep{somov1997,tsuneta1998,tanuma2005}. Recently acceleration from contracting plasmoids \citep{drake2006} and acceleration in plasmoids interacting with fast shocks \citep{nishizuka2013} have been also proposed. It has been argued that a high-density region can be formed as a result of the shock-shock interaction at the top \citep{hori1997} and by the compression at termination shocks \citep{yokoyama2001}. \citet{zenitani2011} carefully analyze shocks in a plasmoid in a ideal 2D MHD simulation. Despite the importance, the shock formation in and above the post-flare loops has not been yet investigated in detail.

How post-flare loops evolve in many astrophysical systems is one of our central interests. Since reconnection produces hot and high-speed plasma flows, it is expected that the thermal structure should be determined by a coupling among the plasma flows, shocks, and heat conduction. The heat conduction is essential to drive the hot high-speed evaporation flows. However, the calculation of the plasma dynamics with the heat conduction effects is computationally expensive. This makes the extensive parameter survey in the multi-dimensional MHD simulations difficult. The development of a simplified post-flare loop model based on a reconnection model is therefore desired. 

In this paper, we study the shock formation and evolution of the thermal structure in and above the post-flare loops using MHD simulations. To understand the shock formation in 2D systems, a 2D MHD simulation of a solar flare has been carried out. We found new shock structures in and above the post-flare loops, which were not well resolved in the previous work by \citet{yokoyama2001}. To study the dynamics of flows along the reconnected magnetic field in detail, kinematics and energetics of the plasma are investigated along selected field lines. It is found that shocks in the post-flare loops are crucial to determine the thermal and flow structures in the post-flare loops. On the basis of the results of the 2D MHD simulation, we have developed a new post-flare loop model which we call the pseudo-2D MHD model. The model is based on the 1D MHD equations, and has been developed to model a 3D reconnected field line. Through a comparison, it is found that the shock formation and thermal evolution in the pseudo-2D MHD model are similar to those in the 2D MHD model.

This paper is structured as follows. In Section~\ref{sec:2dmhd}, the 2D MHD model of a solar flare is introduced. In Section~\ref{sec:2dmhd_result}, numerical results of the 2D MHD simulation are detailed. In Section~\ref{sec:p2dmhd}, on the basis of the 2D MHD simulation, we develop our pseudo-2D MHD model of a post-flare loop. In Section~\ref{sec:p2dmhd_result}, numerical results of the pseudo-2D MHD model are introduced and are compared with the 2D MHD simulation. Section~\ref{sec:discussion} contains our summary and discussion.

\section{2D MHD Model of a Solar Flare}\label{sec:2dmhd}
\subsection{Assumptions and Basic Equations}
We performed a 2D MHD simulation of a solar flare similar to \citet{yokoyama2001} simulations. We take a rectangular calculation domain in the $x$-$y$ plane.  The MHD equations in the following form are solved:
\begin{align}
\frac{\partial \rho}{\partial t} + \nabla \cdot(\rho \bm{v}) &=0,\\
\frac{\partial \rho \bm{v}}{\partial t} + \nabla \cdot \left[ \rho \bm{vv} 
+ \left( p+\frac{|\boldmath{B}|^2}{8\pi}\right) \underline{\bm{1}} - \frac{\bm{BB}}{4\pi} \right] &= 0,\\
\frac{\partial \bm{B}}{\partial t}+c\nabla\times\bm{E} & = 0, \\
\frac{\partial e}{\partial t} + \nabla \cdot \left[ \bm{v} \left( e + p+\frac{|\boldmath{B}|^2}{8\pi}\right) - \frac{1}{4\pi}\bm{B}(\bm{v}\cdot \bm{B}) + \frac{c\eta}{4\pi}\bm{J}\times\bm{B} + \bm{F}_c \right] &=0,\\
\bm{F}_c=-\kappa_{0}T^{5/2}\nabla_{\parallel}T,&\\
p= \frac{k_B}{m}\rho T,&\\
\bm{E}=\eta \bm{J}-\frac{1}{c}\bm{v}\times\bm{B},&\\
e= \frac{p}{\gamma-1}+\frac{1}{2}\rho \bm{v}^2+\frac{\bm{B}^2}{8\pi},&\\
\bm{J}=\frac{c}{4\pi}\nabla\times\bm{B},&
\end{align}
where $\bm{v}=(v_x,v_y,v_z)$ and $\bm{B}=(B_x,B_y,B_z)$. $\kappa_0$ ($\sim 10^{-6}$ in cgs units) is the coefficient of the Spitzer-type conductivity, $\gamma$ is the specific heat ratio (5/3 is used in this study), $m$ is the mean particle mass and $k_B$ is the Boltzmann constant. $\eta$ is the electric resistivity. $\bm{F}_c$ is the conduction flux. The normalization units of our simulations are summarized in Table~\ref{tab:units}. The radiative cooling time is expected to be longer than the dynamical time in the post-flare loops. Since we mainly focus on the dynamical processes such as the shock formation and flows, we neglect the radiative cooling.

\subsection{Initial and Boundary Conditions}
\begin{figure}
\epsscale{.60}
\plotone{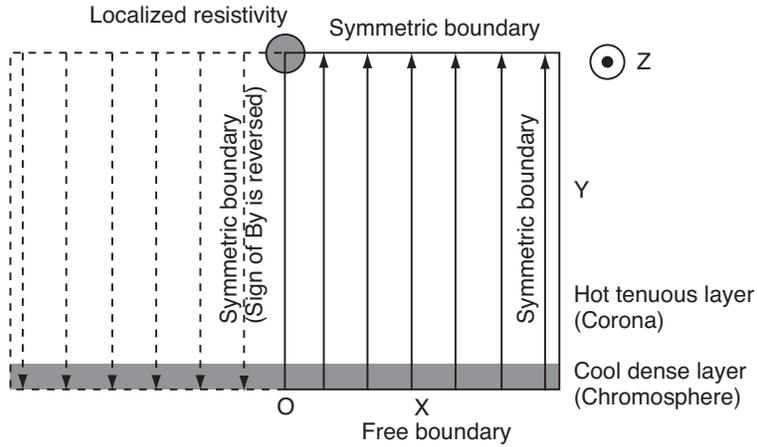}
\caption{Schematic diagram of the initial and boundary conditions of the 2D MHD simulation. The solid arrows denote the magnetic field. The circled region indicate the region where the localized resistivity is applied. The grey region at the bottom is the chromosphere. The calculated domain is the region where $x>0$. \label{fig:ic_2dmhd}}
\end{figure}

Our model is identical to the \citet{yokoyama2001} model, except for the domain size and boundary conditions. The initial condition is same as their model. The initial and boundary conditions are as follows. The domain is $0\le x \le x_{max,2D}$ and $0\le y \le y_{max,2D}$, where $x_{max,2D}$=10 and $y_{max,2D}=20$. A schematic diagram of the initial and boundary conditions are shown in Figure~\ref{fig:ic_2dmhd}. The initial magnetic field is assumed to be a force-free field and is given by
\begin{align}
B_x(x,y) & = 0, \\
B_y(x,y) & = B_0\tanh{(x/w)},\\
B_z(x,y) & = B_0/\cosh{(x/w)},
\end{align}
where $w=0.5$ is the width of the initial current sheet. The gas pressure is assumed to be uniform ($p_0$). The density distribution is described as
\begin{align}
\rho(x,y) & = \rho_{chr} + (\rho_{cor}-\rho_{chr})\frac{1}{2}\left( \tanh{\left[(y-h_{TR})/w_{TR,2D}\right]} +1\right),
\end{align}
where $\rho_{cor}$ and $\rho_{chr}$ are respectively the densities in the corona and chromosphere, and $h_{TR}$ and $w_{TR,2D}$ are respectively the height and the width of the transition region between the corona and chromosphere. $h_{TR}$ and $w_{TR,2D}$ are set to 1 and 0.2, respectively. The chromosphere is modeled as a dense and cool plasma. For simplicity, in this paper $\rho_{chr}$ is set to $10^5 \rho_{cor}$. $T_{cor}=(m/k_B)(p_0/\rho_{cor})$ is the initial coronal temperature.

To allow the magnetic field to reconnect, we impose a localized resistivity in the form of
\begin{align}
\eta(x,y) &= \eta_0 \exp{\left[ -(r/w_\eta)^2 \right]},
\end{align}
where $r=\sqrt{x^2 + (y-h_{TR})^2}$ and $w_\eta=1$. We localize the resistivity and fix it with time to realize a fast and quasi-steady magnetic reconnection with a single X-point \citep{ugai1992}.

All the boundaries are symmetric. At the boundary at $x=0$ the sign of $B_y$ is reversed.

The numerical scheme is based on a Harten-Lax-van Leer (HLL) scheme developed by \citet{miyoshi2005}, HLLD ("D" stands for Discontinuities), which is a shock-capturing scheme. It has the second-order accuracy in space and time. The heat conduction term is solved with an implicit method similar to \citet{yokoyama2001} method to reduce the calculation time. We modify their method to more accurately calculate the heat conduction flux. A detailed description of the implicit method is given in Appendix~\ref{ap:A}. The calculated domain is resolved with $800\times1000$ grids.

\section{Numerical Results of 2D MHD Model}\label{sec:2dmhd_result}
\subsection{Overview of Evolution}

\begin{figure}
\epsscale{.90}
\plotone{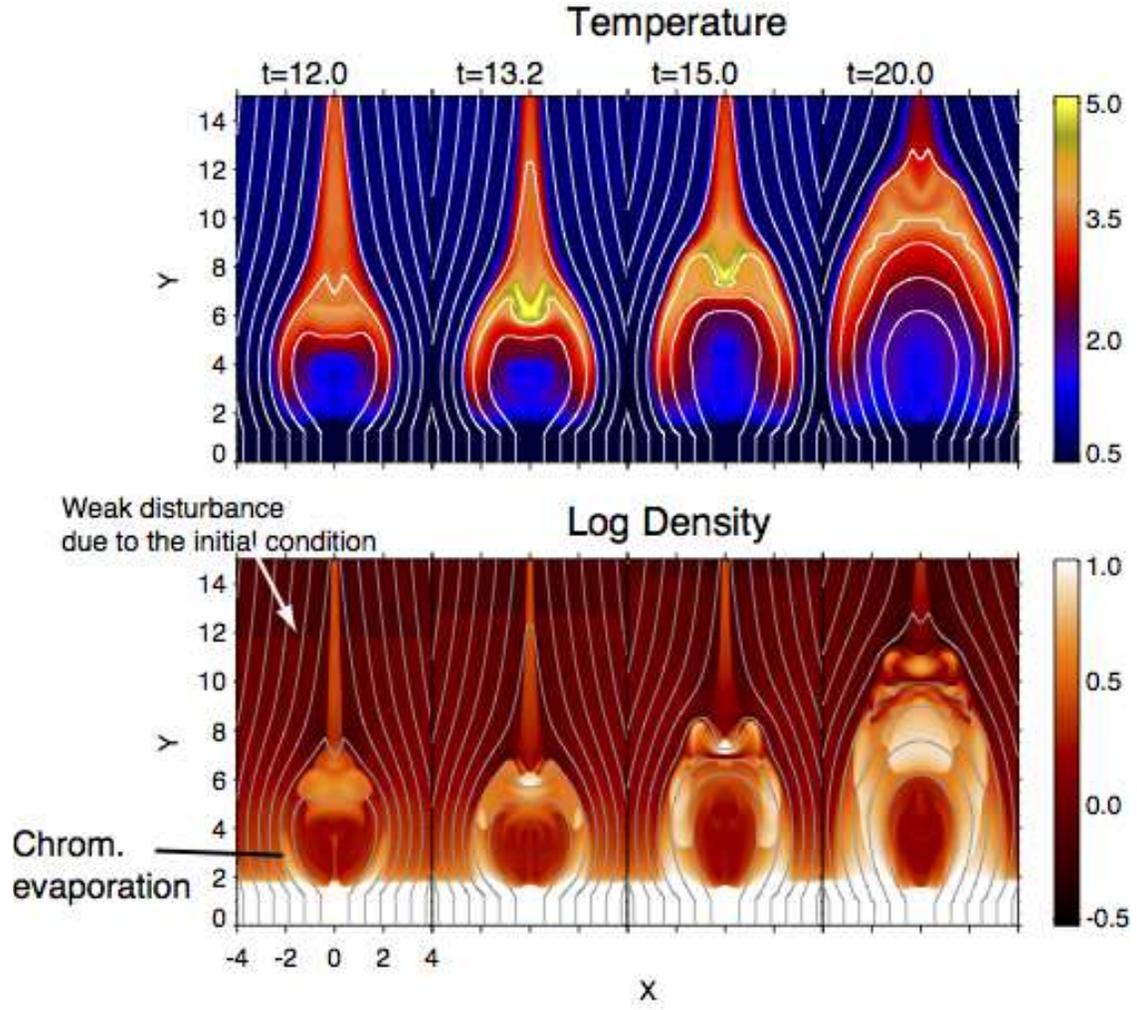}
\caption{Time evolution of the temperature (Top) and density (Bottom) of the 2D MHD model. The solid lines denote the magnetic field.\label{fig:overview_2dmhd}}
\end{figure}

Figure~\ref{fig:overview_2dmhd} displays an overview of the time evolution of the 2D MHD simulation. The region where $x < 0$ is also shown only for visual inspection. Reconnection takes place due to the localized resistivity and the Alfvenic collimated reconnection outflow is produced. The reconnected field lines pile up in the outflow region to form the growing loop system at the base of the corona. The temperature structure is smooth along the magnetic field owing to the heat conduction. The heat in the hot outflow is carried to the chromosphere along the magnetic field due to the heat conduction. As a result, the upper chromosphere is heated up and rapidly expands, leading to the formation of the hot dense upflows called the chromospheric evaporation. The evaporated plasma finally fills the reconnected field lines and forms the hot dense post-flare loops. We note that the weak disturbance which starts to propagate at the beginning of the simulation is generated because the initial condition is not in the thermal equilibrium between the chromosphere and corona. We confirmed that its effect is negligible for the dynamics of the post-flare loops.

\begin{figure}
\epsscale{.80}
\plotone{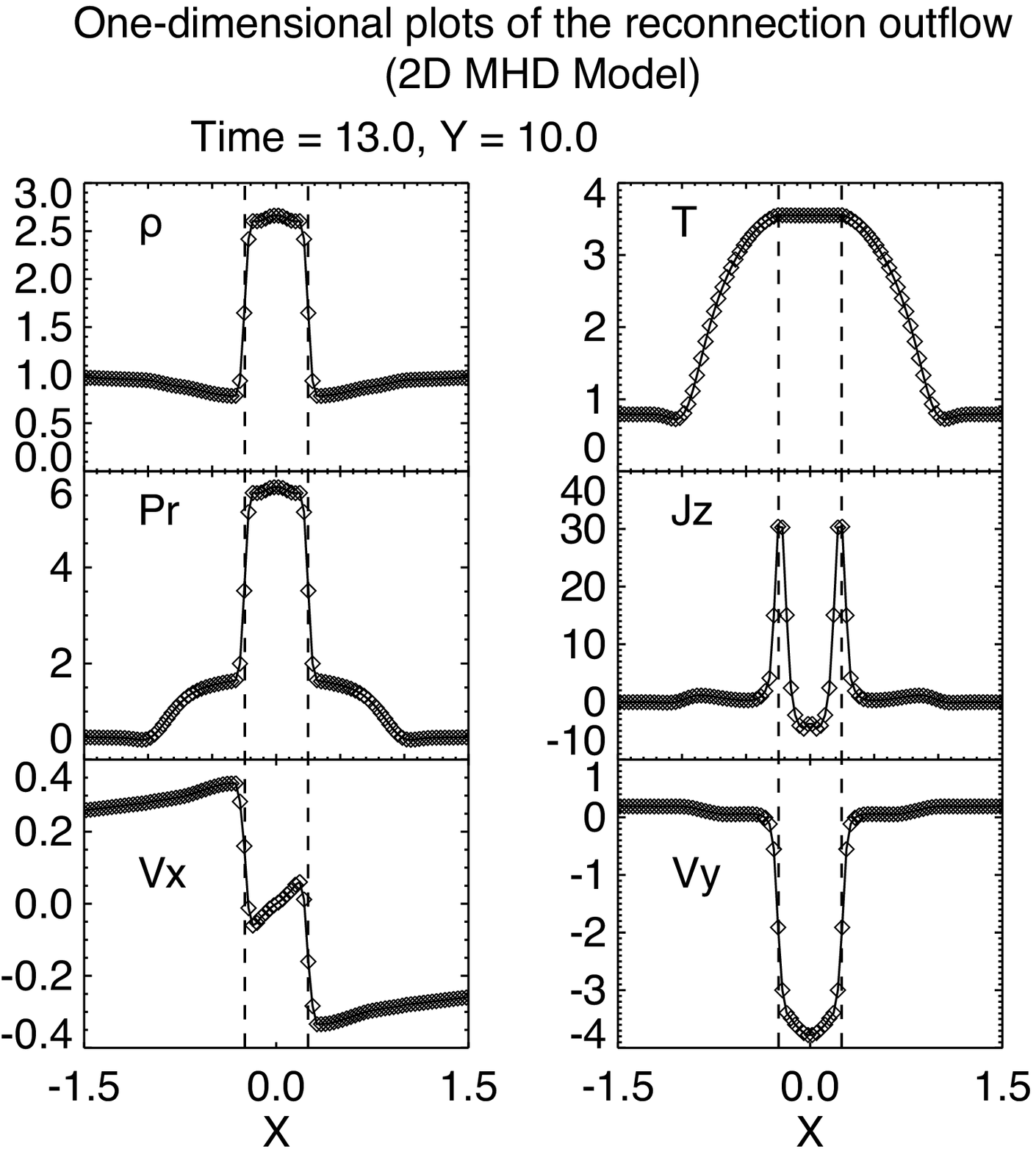}
\caption{One-dimensional plots parallel to the $x$-axis across $y=10$ at $t=13$ (across the reconnection outflow region). The density, temperature, pressure, $J_z$, $v_x$ and $v_y$ are displayed. The vertical dashed-lines indicate the slow shocks attached to the reconnection outflow. \label{fig:shock_xcut_2dmhd}}
\end{figure}

One-dimensional plots parallel to the $x-$axis across $y=10$ (i.e. across the reconnection outflow) are shown in Figure~\ref{fig:shock_xcut_2dmhd}. It is shown that a Petschek-type reconnection is established because of the localized resistivity, where two slow shocks emanate from the reconnection region. The slow shocks can be discerned as a pair of the discontinuities in Figure~\ref{fig:shock_xcut_2dmhd}. We note that due to the heat conduction the slow shocks become isothermal slow shocks \citep{forbes1989,yokoyama1997,chen1999,seaton2009}.

\begin{figure}
\epsscale{.90}
\plotone{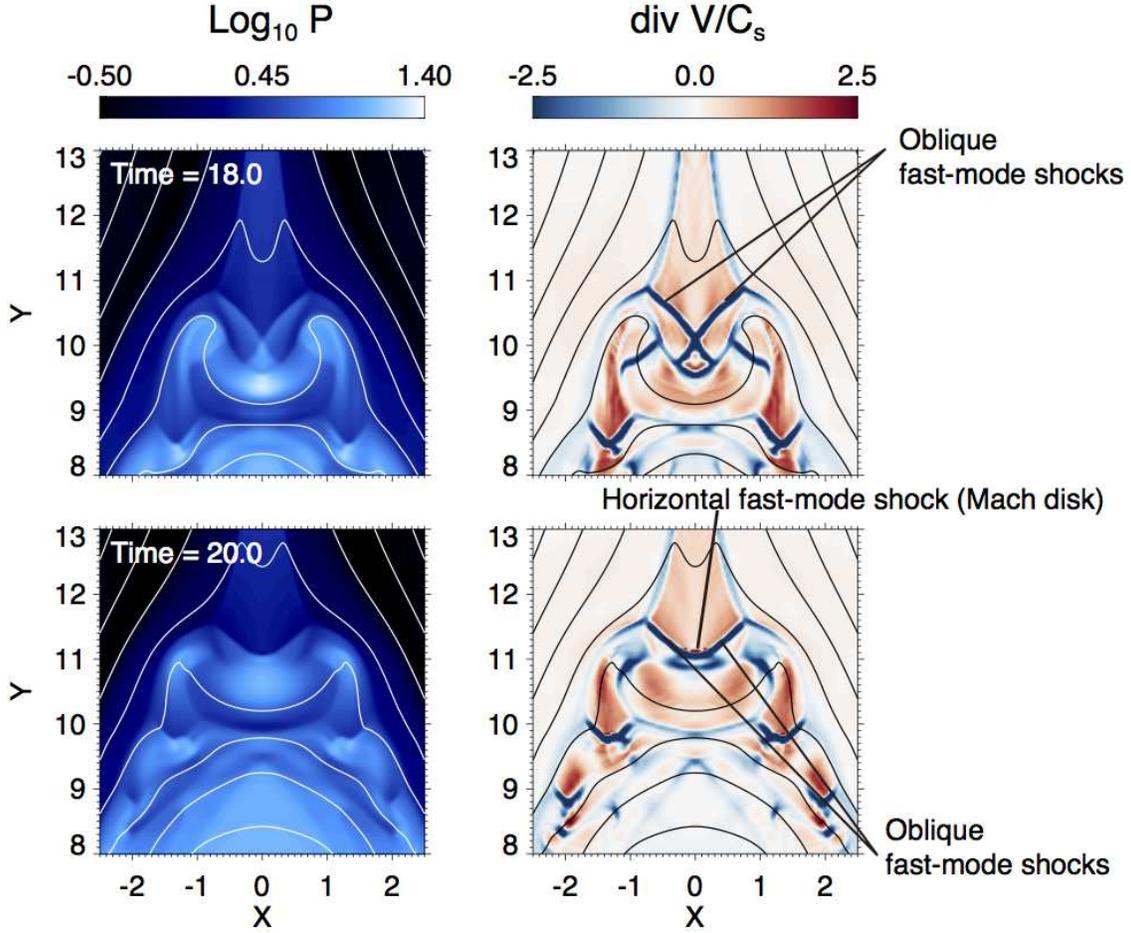}
\caption{Gas pressure (Left) and $\nabla \cdot \bm{v}/C_s$ (Right) maps of the loop top. Top: Snapshot when the termination shock consists of two oblique fast-mode shocks. Bottom: Snapshot when the termination shock consists of two oblique shocks and a horizontal shock. Note that most of the regions where $\nabla \cdot \bm{v}/C_s$ takes large negative values are slow or fast shocks. \label{fig:pr_top}}
\end{figure}

The thermal structure of the post-flare loops is determined by a complex coupling among the plasma flows, shocks, and heat conduction. When the reconnection outflow collides with the loop system below, two oblique fast-mode shocks and sometimes a horizontal fast-mode shock (Mach disk) are formed at the top (see Figure~\ref{fig:pr_top}), which indicates that the termination shock is a combination of two or three fast-mode shocks. Note that most of the regions where $\nabla \cdot \bm{v}/C_s$ takes large negative values are slow or fast shocks. We found that the strength of the termination shock shows a quasi-periodic oscillation. We also found the shock reflection and Mach reflection in a concave magnetic field at the top (Figures~\ref{fig:pr_top} and \ref{fig:ro_looptop_2dmhd}). The relationship between the oscillation and the flow structure at the top will be discussed in our future papers. 

The high-pressure plasma at the loop-top expands along the magnetic field to the foot-points, generating the high-speed downflows. Then the evaporation flows collide with the downflows, forming the dense regions in the post-flare loops ("humps" in Figure~\ref{fig:ro_hump_2dmhd}, named by \citet{yokoyama2001}). After the collision, the fronts of the evaporation flows and downflows becomes slow shocks (see $\nabla \cdot \bm{v}/C_s$ maps). The pair of the upward slow shocks finally collide with each other at the apex, forming another dense region (see Figure~\ref{fig:ro_looptop_2dmhd}). Note that the "blob" named by \citet{yokoyama2001} is different from this high-density region. The blob is the high-density region in a concave magnetic field at the top. 

Figure~\ref{fig:ro_te_ent_beta_2dmhd} displays the slow shocks in the post-flare loops mentioned above. The entropy is discontinuous at the two pairs of the discontinuities in the density map, but the temperature is smooth along the magnetic field, indicating that they are isothermal slow shocks. The plasma $\beta$ in a large fraction of the post-flare loops, as well as in the outflow region, is larger than unity (the contour indicates the level where $\beta=1$), meaning that the low-$\beta$ approximation cannot be applied to the reconnected magnetic field. This is a consequence of the shock heating and compression. The high-pressure post-flare loops are confined by the surrounding low-$\beta$ plasma. This is consistent with the observation by \citet{mckenzie2013}; they concluded that plasma $\beta$ is of the order of unity or larger than unity in the supra-arcade plasma in two flares analyzed.

\subsection{Dynamics and Energetics along a Specific Field Line}\label{subsec:line_2dmhd}
We have seen the two-dimensional evolution of a solar flare. To study the dynamics of flows along the reconnected magnetic field in detail, kinematics and energetics of the plasma are investigated along selected field lines. We performed the same analysis for other field lines and confirmed that they give similar results.

We picked up a magnetic field line and measured the physical quantities along it. The field line used in the analysis is shown in Figure~\ref{fig:ro_line_2dmhd}. Figure~\ref{fig:ro_te_vp_divv_2dmhd} displays the time-distance diagrams of the density, temperature, $v_{\parallel}=|(\bm{v}\cdot\bm{B}) \bm{B}/B^2|$ and $\nabla\cdot \bm{v}/C_s$ along the field line whose foot-point is located at $(x,y)=(1.2,0)$, where $C_s$ is the local sound speed. The distance is measured along the field line. Before the field line reconnects, the starting point of the distance is the intersection point between the field line and the top boundary. After reconnection, the starting point is the apex of the closed loop. The sign of $v_{\parallel}$ is defined as positive and negative when a plasma flow travels to the apex and foot-points, respectively.

In Figure~\ref{fig:ro_te_vp_divv_2dmhd}, we can discern the field line shrinking after reconnection. The heat released at the Petschek slow shocks is transferred to the chromosphere (see the temperature in Figure~\ref{fig:ro_te_vp_divv_2dmhd}), leading to the generation of the chromospheric evaporation. The evaporation flows are seen as upflows from the chromosphere (see $\rho$ and $v_{\parallel}$ maps). The reconnection outflow is significantly decelerated at $t\sim12$. This is seen as the enhancement of the density and temperature and negative $\nabla \cdot \bm{v}$ (compression), indicating that the kinetic energy of the outflow is converted to the thermal energy. After the termination, the high-pressure plasma at the top expands along the magnetic field, forming the downflows. The downflows collide with the chromospheric evaporation flows, leading to the formation of the humps (see also Figure~\ref{fig:ro_hump_2dmhd}). After the collision, the fronts of the evaporation flows and downflows become steep and finally become slow shocks. The pair of the upcoming shocks finally cross each other at the apex at $t\sim16$, forming another high-density region (see also Figure~\ref{fig:ro_looptop_2dmhd}).

We found slow shocks propagating along the field line, but there is no prominent shocks nor waves propagating back and forth from end to end. They are damped by the heat conduction \citep[e.g.][]{ofman2002}. Also note that the propagation speed of the shocks is largely decelerated by the evaporation flow (i.e. Doppler effect), which significantly affects the traveling time of the slow-mode waves/shocks (the local sound speed is $\sim1.5 $, but the propagation speed is $\sim 0.6$). That is, a simple estimation by $t_{travel,slow}\sim L/C_s$ is not a good approximation of the traveling time, where $L$ and $C_s$ is the loop length and the sound speed in the loop, respectively. We observe no prominent signature of the standing slow-mode waves in the calculated time range (as for acoustic waves in the post-flare loops, see e.g. \citet{nakariakov2004}).

Figure~\ref{fig:energy_line_2dmhd} shows the time evolution of the total, magnetic, internal (thermal) and kinetic energies integrated along two specific field lines. The top and bottom panels are for the field lines which are rooted at $(x,y)=(1.2,0)$ and $(1.8,0)$, respectively. Considering that the cross-sectional area of the flux tube is inversely proportional to the magnetic field strength, we integrate the energies along a field line as follows:
\begin{align}
E_{mag} &=  \int ds \frac{B^2}{8\pi} \frac{1}{B},\\
E_{int} &=  \int ds \frac{p}{\gamma-1} \frac{1}{B},\\
E_{kin} &=  \int ds \rho\frac{v^2}{2} \frac{1}{B},\\
E_{tot} & = E_{mag}+E_{int} + E_{kin}.
\end{align}
In this paper, all the energies are normalized by the initial total energy.
After the field lines reconnect ($t\sim 8$), the magnetic energy is rapidly converted to the internal and kinetic energies. As for the field line rooted at $(x,y)=(1.2,0)$, the maximum value of the kinetic energy is 0.35. Let us define the time when $E_{kin}$ becomes $E_{kin,max}$ as $t_{peak}$. If we compare the total, magnetic and internal energies at $t=0$ and those at $t=t_{peak}+10$, $dE_{tot}=E_{tot}(t=t_{peak}+10)-E_{tot}(t=0)\sim0.15$, $-dE_{mag}=-(E_{mag}(t=t_{peak}+10)-E_{mag}(t=0))\sim0.45$, and $dE_{int}=E_{int}(t=t_{peak}+10)-E_{int}(t=0)\sim0.6$. Note that the variation of the total energy remains within $\sim 15$~\%. The field line rooted at $(x,y)=(1.8,0)$ also gives a similar result. The small variation in the total energy reflects the fact that the compressive and expanding motions by the surrounding plasma are localized in time and space (see Figure~\ref{fig:ro_te_vp_divv_2dmhd}). Therefore, the compressive motion of the reconnected flux tube is found to be not significant with respect to the total energy variation.

\section{Pseudo-2D MHD Model of a Post-Flare Loop}\label{sec:p2dmhd}
\subsection{Physical Processes Considered}
Through the analysis of the 2D MHD simulation, we found that the field-aligned plasma motions (particularly evaporation flows and slow shocks) and heat conduction seem to mainly determine the dynamics in the post-flare loops. Fast shocks are important for converting the kinetic energy of the reconnection outflow to the heat and for locally changing the cross-sectional area, but they do not largely change the total energy of the field lines (Figure~\ref{fig:energy_line_2dmhd}). On the basis of the results, we aim to model the multi-dimensional reconnection and flare processes in a simplified MHD scheme. 

What is important for conduction-heating-type flare modeling would be 1. to model a reconnected field line, 2. to model the reconnection inflow and outflow, 3. to include MHD waves (since MHD waves can carry a large amount of the released energy from the reconnection sites \citep{kigure2010}), 4. termination of the reconnection outflow and energy conversion of the kinetic energy of the outflow into the thermal energy, and 5. to include the heat conduction which is essential to determine the temperature of the flare loops.

Considering that the plasma motions are frozen-in a magnetic field, a 1D MHD model will be the simplest form among the possible MHD models. We developed a model based on the 1D MHD equations, where all the variables depend on one space dimension and all the three components of the magnetic and velocity fields are considered. We regard our model as a pseudo-2D MHD model. Note that the meaning of "pseudo-2D" of our model is different from that of \citet{hori1997} hydrodynamic model which consists of isolated and fixed semi-circular loops with different lengths and constant cross-section.

\subsection{Assumptions and Basic Equations}
We take a cartesian coordinate system in which all the variables are functions of $x$. The reconnection outflow is in the $y$-direction. $z$ is the ignorable coordinate (uniform in the $z$-direction). 

We mimic a reconnected magnetic field line by assuming a magnetic field with a sharp bend (see Figure~\ref{fig:1d_concept}). The magnetic field line shrinks with time and drives the flow perpendicular to the field which represents the reconnection outflow. 

A schematic picture of our model is as follows. Figure~\ref{fig:2d_vs_1d} displays a comparison of our pseudo-2D MHD model with the 2D MHD model. $h_{rx}$ is the height of the reconnection point, and $x_{max}$ is the location of the foot-point of the field line. They are linked by the relation of $x_{max}=h_{rx}\tan{\theta}$. The reconnection angle $\theta$ and the initial plasma beta $\beta$ are treated as free parameters. If a guide-field (the $z$-component of the magnetic field) is included, the guide-field angle $\phi=\tan^{-1}{(B_z/B_x)}$ will be an additional parameter.

The shrinking motion of the reconnected field line will stop when it collides with the magnetic loops piling up below. To model this process, the final configuration of the field line is given in the model and the outflow is decelerated when the field line approaches the final state. The termination process is modeled by adding a damping term to the equation of motion perpendicular to the moving field line. We let the damping term work only when the reconnected field line comes close to the final configuration. 

According to Figure~\ref{fig:energy_line_2dmhd}, the total energy of a field line is conserved within several 10~\%. On the basis of this result, we hypothesize that the total energy in a reconnected flux tube is conserved. The cross-sectional area in our model is assumed to be uniform and constant in time.

The basic equations are as follows:
\begin{align}
\frac{\partial \rho}{\partial t} + \frac{\partial}{\partial x}(\rho v_x)& = 0,\\
\frac{\partial B_y}{\partial t} -\frac{\partial }{\partial x} (v_y B_x - v_x B_y ) &= 0,\\
\frac{\partial B_z}{\partial t} -\frac{\partial }{\partial x} (v_z B_x - v_x B_z ) &= 0,\\
\frac{\partial e}{\partial t} 
+ \frac{\partial}{\partial x}\left[ (e + p + \frac{|\bm{B}|^2}{8\pi})v_x - \frac{1}{4\pi}B_x(\bm{v}\cdot\bm{B}) 
+ \kappa_0 T^{5/2} \frac{B_x^2}{|\bm{B}|^2}\frac{\partial T}{\partial x}\right] & = 0,\\
\frac{\partial (\rho v_x) }{\partial t} 
+ \frac{\partial}{\partial x} \left[ \rho v_x^2 + p + \frac{|\bm{B}|^2}{8\pi} - \frac{B_x^2}{4\pi}\right]& 
= -\nu_b(\Delta d(x,t))\rho v_{\perp x},\\
\frac{\partial (\rho v_y) }{\partial t} 
+ \frac{\partial}{\partial x} \left[ \rho v_x v_y  - \frac{B_xB_y}{4\pi}\right]& 
= -\nu_b(\Delta d(x,t))\rho v_{\perp y},\\
\frac{\partial (\rho v_z) }{\partial t} 
+ \frac{\partial}{\partial x} \left[ \rho v_x v_z  - \frac{B_xB_z}{4\pi}\right]& 
= 0,\\
e = \frac{p}{\gamma-1} + \frac{1}{2}\rho |\bm{v}|^2 + \frac{|\bm{B}|^2}{8\pi}, &\\
p = \frac{k_B}{m}\rho T,&
\end{align}
where $\bm{v}=(v_x,v_y,v_z)$, $\bm{B}=(B_x,B_y,B_z)$, $v_{\perp x}=(\bm{v}-\bm{v}_\parallel)_x$, $v_{\perp y}=(\bm{v}-\bm{v}_\parallel)_y$, and $\bm{v}_\parallel=\bm{B}(\bm{v}\cdot\bm{B})/|\bm{B}|^2$. All the physical quantities are only dependent of $x$ and $t$. Note that we include a damping term in the momentum equations that slows down only the x and y-components of the velocity perpendicular to the magnetic field. The detailed description of the damping term is given below. Note that the total energy is conserved along a field line, and that the kinetic energy reduced by the damping term is converted into the thermal energy. The normalization units are the same as those in the 2D MHD model (See Table~\ref{tab:units}). Note that this model can treat slow-mode, fast-mode and Alfven waves.

Considering the symmetry, we only solve the domain within $0\le x \le x_{max}$. At the boundary of $x=0$ the mirror symmetric boundary conditions are imposed. The boundary of $x=x_{max}$ is free.

The initial conditions are as follows. The free parameters that determine the initial magnetic field are the plasma beta $\beta=8\pi p/|\bm{B}|^2$, and the angles $\theta$ and $\phi$ (see Figure~\ref{fig:2d_vs_1d}). The guide field effect is detailed in Appendix~\ref{ap:B}. The initial magnetic field is given by
\begin{align}
B_0 & = \sqrt{\frac{8\pi p_0}{\beta}},\\
B_x(x) & = -B_0 \sin{\theta} \cos{\phi},\\
B_y(x) & =B_0 \cos{\theta} \cos{\phi},\\
B_z(x) & = B_0 \sin{\theta} \sin{\phi}.
\end{align}
The gas pressure is assumed to be uniform ($p_0$).
The domain is divided into the two layers, namely chromosphere and corona:
\begin{equation}
\rho(x) = \rho_{cor} + (\rho_{chr}-\rho_{cor})\frac{1}{2}\left[ 1+ \tanh{ \left( \frac{x-x_{TR}}{w_{TR}} \right)}\right],
\end{equation} 
where $x_{TR}$ is the location so that $y_{line}(x_{TR},t=0)=h_{TR}$. Where $y_{line}(x,t)$ is a function that describes the configuration of the field line.

The initial and final magnetic field configurations are described as follows. See also Figure~\ref{fig:def_deltad}. The reconnection point (the location of the sharp bend) is assumed to be at $(x,y)=(0,h_{rx})$. The initial condition of a magnetic field is written as
\begin{equation}
y_{line}(x,t=0) = -\frac{1}{\tan{\theta}}(x-x_{max})\label{eq:initial_mag},
\end{equation}
for $x\ge0$. The final state of the magnetic field is approximated by a quadratic function of
\begin{equation}
y_{line,fin}(x) = -\frac{x_{max}}{2\tan{\theta}}\left[\left(\frac{x}{x_{max}}\right)^2-1\right],
\end{equation}
which has the same slope of the tangent line with the equation~(\ref{eq:initial_mag}) at $x=x_{max}$. 

We virtually consider the travel distance of the reconnected field line in the $y$-direction. When the field line approaches the final state, the damping term is applied only to the $x$- and $y$-components of the velocity perpendicular to the field line. We define the distance in the $y$-direction between the field line at $(x,t)$ and the field line in the final state as
\begin{equation}
\Delta d(x,t) = y_{line}(x,t)-y_{line,fin}(x).
\end{equation}
The damping term only works when the field line approaches to the final state:
\begin{equation}
\nu_d(\Delta d(x,t)) = \frac{1}{t_{damp}}\frac{1}{2}\left[ 1-  \tanh{\left(\frac{\Delta d(x,t)}{w_d}\right)} \right],
\end{equation}
where $t_{damp}=w_d/v_{A,y}$, $V_{A,y}=B_y/\sqrt{4\pi\rho_{cor}}$ is the outflow speed in the y-direction, and $w_d$ is a free parameter that denotes a typical braking distance. To prevent the field line from shrinking further even after the travel time $t_{travel}=(h_{rx}-y_{line,fin}(x=0))/V_{A,y}$, we increase the damping coefficient $\nu_d$ by a factor of 100 after the time $t=1.2t_{travel}$.
Note that the term arising from the damping terms is not included in the energy equation under the assumption that the total energy along a field line is conserved. The kinetic energy decreased by the damping term is converted only into the thermal energy.

The reconnection angle $\theta$ and plasma beta $\beta$ are important parameters to determine the total released magnetic energy and energy conversion rate (say, reconnection rate). To choose a physically acceptable parameter set, we utilize the analytical approach by \citet{falle1998}. A detailed description is presented in Appendix~\ref{ap:B}.

The height of the reconnection point $h_{rx}$ is assumed to be $20=y_{max,2D}$, and the reconnection angle $\theta$ is $\pi/12$. The domain size is therefore $x_{max}=h_{rx}\tan{\theta}=20\tan{(\pi/12)}$. The width of the transition region and the typical damping distance are respectively $w_{TR}=w_{TR,2D}=0.2$ and $w_{d}=2$.

The numerical scheme of the pseudo-2D MHD model is based on the \citet{vogler2005}: the fourth-order space-centered difference for spatial derivative and an explicit four-step Runge-Kutta time integration. We explicitly solve the heat conduction term. Using the current computational resources, it is not difficult to explicitly solve the heat conduction in our 1D calculations. The domain is resolved by 640 grids. 

\section{Numerical Results of Pseudo-2D MHD Model}\label{sec:p2dmhd_result}
\subsection{Dynamics and Energetics}
Figure~\ref{fig:line_evo_p2dmhd} demonstrates the time evolution of the magnetic field structure. The field line retracts and sweeps up the plasma like a slingshot. The reconnection outflow is decelerated when the field line approaches the assumed final configuration. In the following, we performed the same analysis as for the 2D MHD model.

One-dimensional plots shown in Figure~\ref{fig:shock_xcut_p2dmhd} demonstrate the formation of a pair of two slow shocks attached to the reconnection outflow, that is, the Petschek-type slow shocks. As well as the 2D MHD model (Figure~\ref{fig:shock_xcut_2dmhd}), the Petschek-type shocks are isothermal shocks due to the heat conduction.

Figure~\ref{fig:ro_divv_line_compare} compares the field-aligned motions in the pseudo-2D MHD model and 2D MHD model. As well as in the 2D MHD model, the heat generated at the Petschek slow shocks is transferred to the chromosphere by the heat conduction, forming the chromospheric evaporation. The humps and high-density regions at the top are found to be formed in the same way as in the 2D MHD model: the downflow-evaporation collision and shock-shock interaction, respectively. The damping of the slow shocks are also observed.

The energy evolution is examined in the pseudo-2D MHD model. When we integrate the energies, we drop off the term $1/B$ in the integrand because in the pseudo-2D MHD model the variation of the cross-sectional area is not considered. Figure~\ref{fig:energy_line_p2dmhd} is the same as Figure~\ref{fig:energy_line_2dmhd} but for the pseudo-2D MHD model. It is shown that $-dE_{mag}\sim0.6$, $dE_{int}\sim 0.55$, and $E_{kin,max}\sim 0.25$, which is similar to the results of the 2D MHD model.

\subsection{Dependence on Parameters}
Previously a formula that determines the post-flare loop temperature was derived under the assumption that the energy input to a loop balances with the conduction cooling rate \citep{fisher1990}. The formula is given by
\begin{align}
T \sim \left( \frac{QL_{loop}^2}{2\kappa_0}\right)^{2/7},
\end{align}
where $Q$ is the volumetric heating rate and $L_{loop}$ is the half length of the magnetic field line of a post-flare loop. The heating rate by magnetic reconnection is determined by the Poynting flux: $Q=B^2/(4\pi)\times V_{A}/L_{loop}$. Using this, the temperature can be written as
\begin{align}
T\propto \beta^{-3/7}h_{rx}^{2/7}\kappa_0^{-2/7},
\end{align}
by assuming that $L_{loop}\sim h_{rx}$. This scaling law was derived by \citet{yokoyama1998,yokoyama2001}. We checked whether the scaling law based on a magnetic reconnection model holds in our pseudo-2D MHD model. 

Figure~\ref{fig:scaling} shows the numerically-obtained $\beta$-$T$, $h_{rx}$-$T$ and $\kappa_0$-$T$ relations. $T_{jet}$ denotes the temperature in the reconnection outflow. $T^{*}$ denotes the maximum temperature after the pair of the two slow shocks generated by the chromospheric evaporation flows collides at the apex.
As shown in Figure~\ref{fig:scaling}, it is found that the temperature in the loop obeys the scaling law well. 

\section{Summary and Discussion}\label{sec:discussion}
In this paper we investigated the flow structure, shock formation and thermal evolution in the post-flare loops using MHD simulations. On the basis of the 2D simulation result, we have developed a new post-flare loop model (the pseudo-2D MHD model). We compare the flow structure, shock formation and energetics of a specific field line in the 2D calculation with those in the pseudo-2D MHD model, and then we find that they give similar results. Here we summarize the results and compare our results with previous studies.

\begin{figure}
\epsscale{0.9}
\plotone{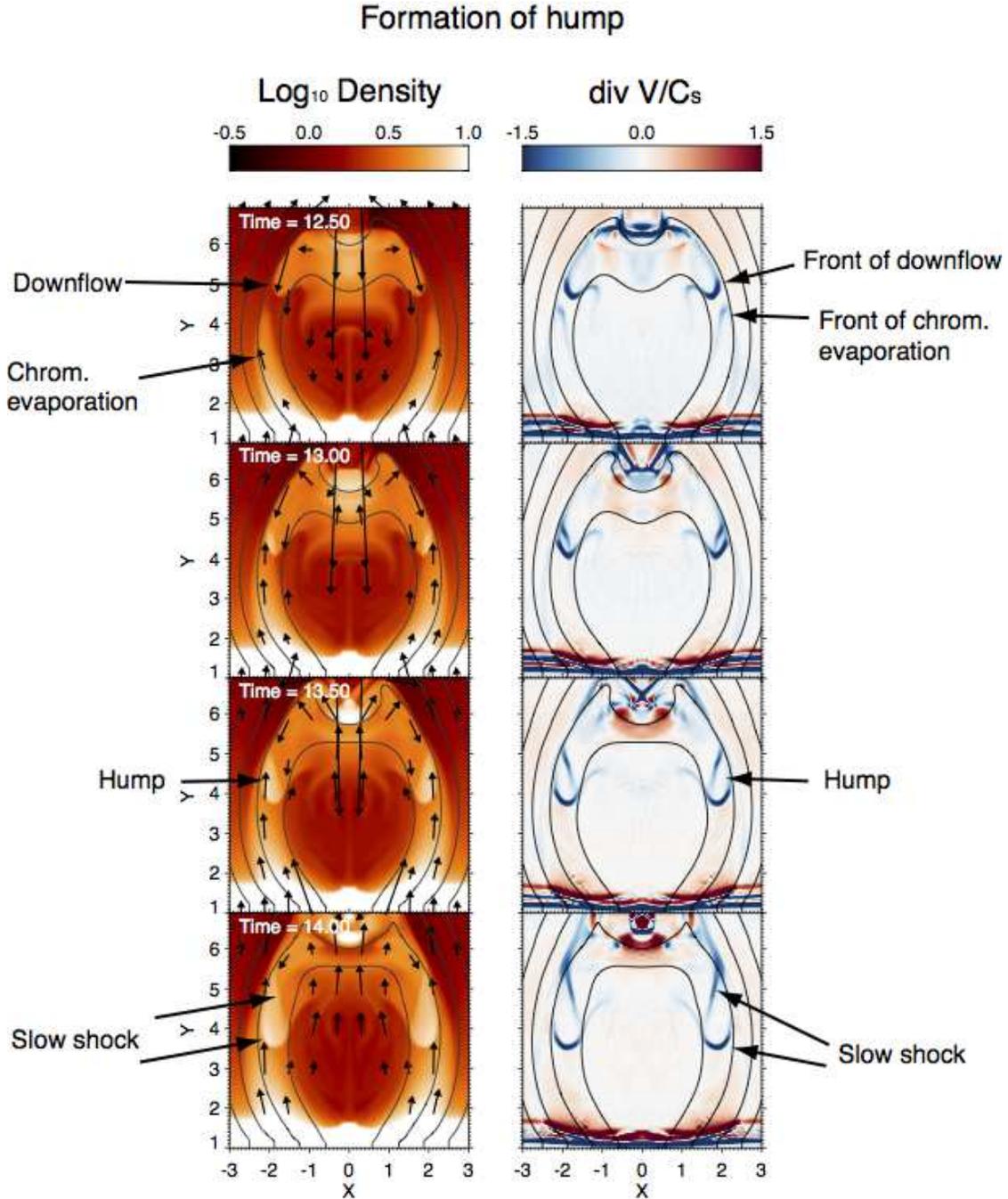}
\caption{Formation of the high-density regions called humps. Left: the density in logarithmic scale. Right: $\nabla\cdot \bm{v}/C_s$. The solid lines denote magnetic field lines. The velocity vectors projected to the x-y plane are also displayed in the density maps (only the vectors with $\sqrt{v_x^2+v_y^2}>0.3$ are shown). Note that most of the regions where $\nabla \cdot \bm{v}/C_s$ takes large negative values are slow or fast shocks. \label{fig:ro_hump_2dmhd}}
\end{figure}

\begin{figure}
\epsscale{1.0}
\plotone{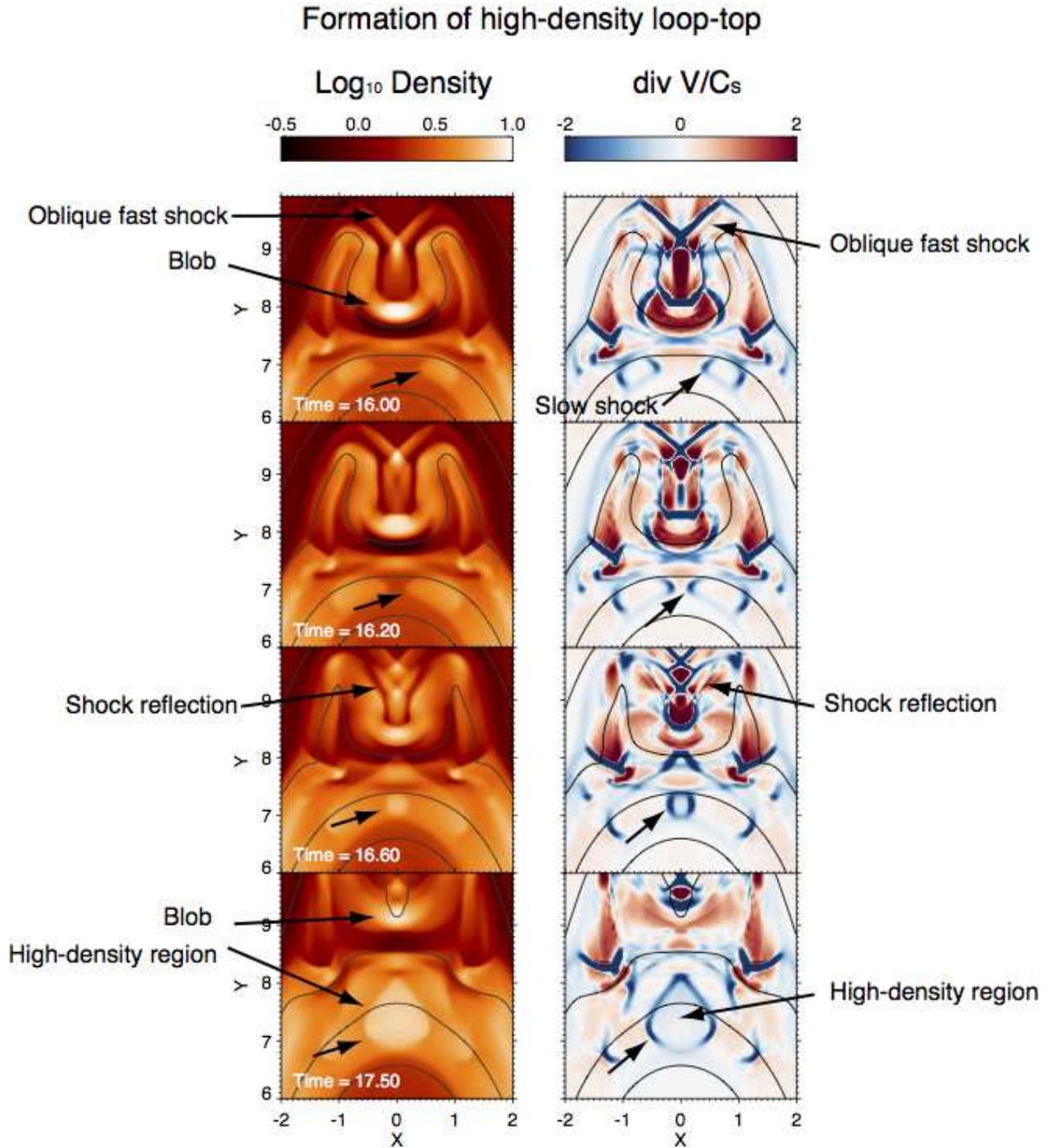}
\caption{Formation of the high-density loop-top. Left: the density in logarithmic scale. Right: $\nabla\cdot \bm{v}/C_s$. The solid lines denote magnetic field lines. Only one of the pair of the slow shocks heading to the apex is indicated by the arrows. Note that most of the regions where $\nabla \cdot \bm{v}/C_s$ takes large negative values are slow or fast shocks. \label{fig:ro_looptop_2dmhd}}
\end{figure}

\begin{figure}
\epsscale{0.7}
\plotone{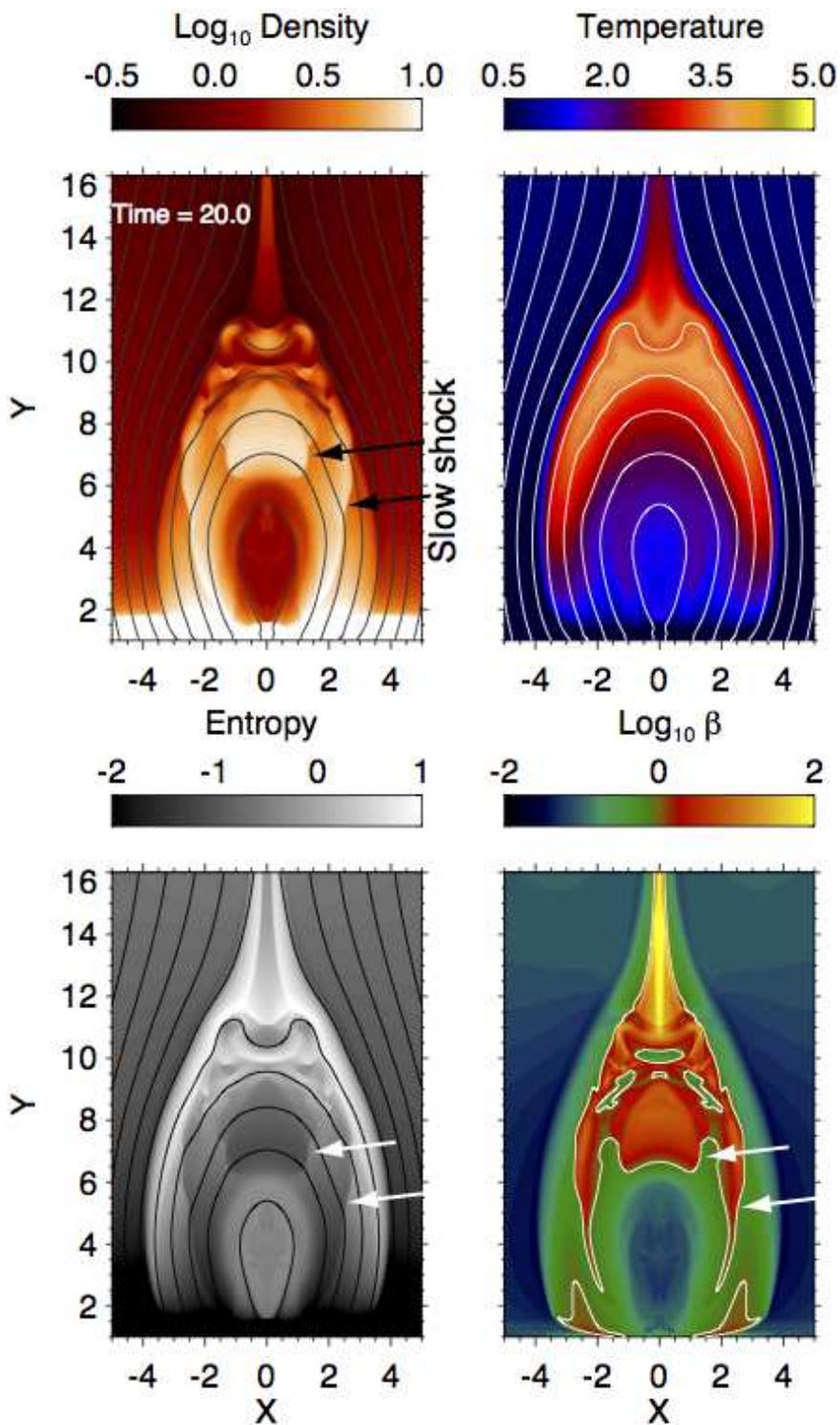}
\caption{Snapshots of the density, temperature, entropy ($\log{(p/\rho^{\gamma})}$), and plasma $\beta$ at $t=20$. The arrows indicate slow shocks in the post-flare loops (See also Figures~\ref{fig:ro_hump_2dmhd} and \ref{fig:ro_looptop_2dmhd}). The solid lines in the first three panels denote the magnetic field. The contour in the plasma $\beta$ map denotes the level where $\beta=1$ ($\log_{10}\beta=0$). \label{fig:ro_te_ent_beta_2dmhd}}
\end{figure}

\begin{figure}
\epsscale{	1.0}
\plotone{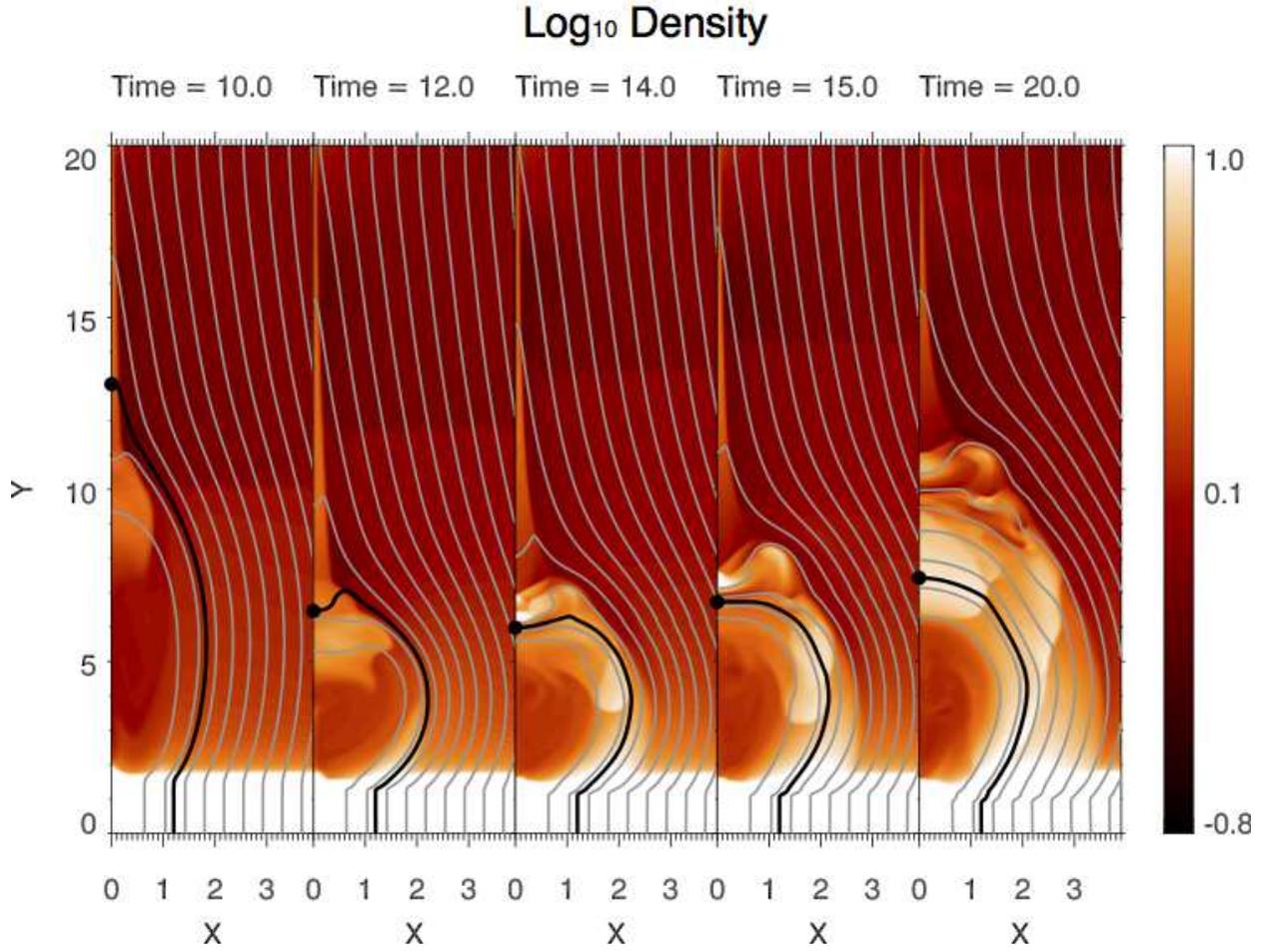}
\caption{Evolution of a specific reconnected field line with the density map. The tracked field line is indicated by the thick black line. The foot-point of the field line is located at $(x,y)=(1.2,0)$. The circle is the origin of the distance. \label{fig:ro_line_2dmhd}}
\end{figure}

\begin{figure}
\epsscale{1.10}
\plotone{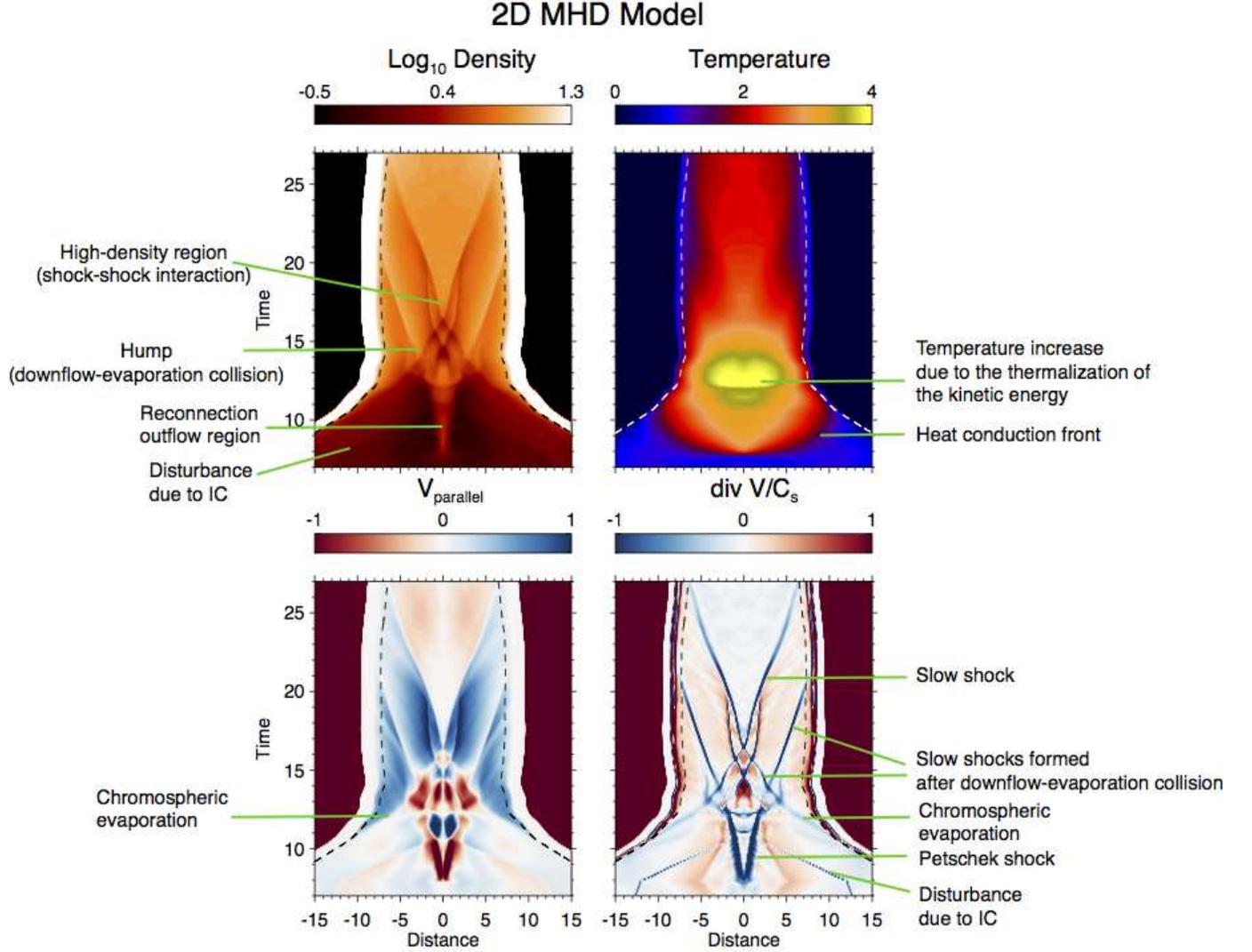}
\caption{Time-distance diagrams of the density (Top Left), temperature (Top Right), $v_{\parallel}$ (Bottom left), and $\nabla\cdot \bm{v}/C_s$ (Bottom Right). The physical quantities are measured along the field line which is originated from $(x,y)=(1.2,0)$. The distance is measured along the field line, and its origin is the apex ($x=0$). The data of the domain of $x<0$ is also shown just for visual inspection. The sign of $v_{\parallel}$ is defined as positive and negative when a plasma flow travels to the apex and foot-points, respectively. This field line does not pass through the termination shock. Note that the regions where $\nabla \cdot \bm{v}/C_s$ takes large negative values are slow or fast shocks.\label{fig:ro_te_vp_divv_2dmhd}}
\end{figure}

\begin{figure}
\epsscale{0.7}
\plotone{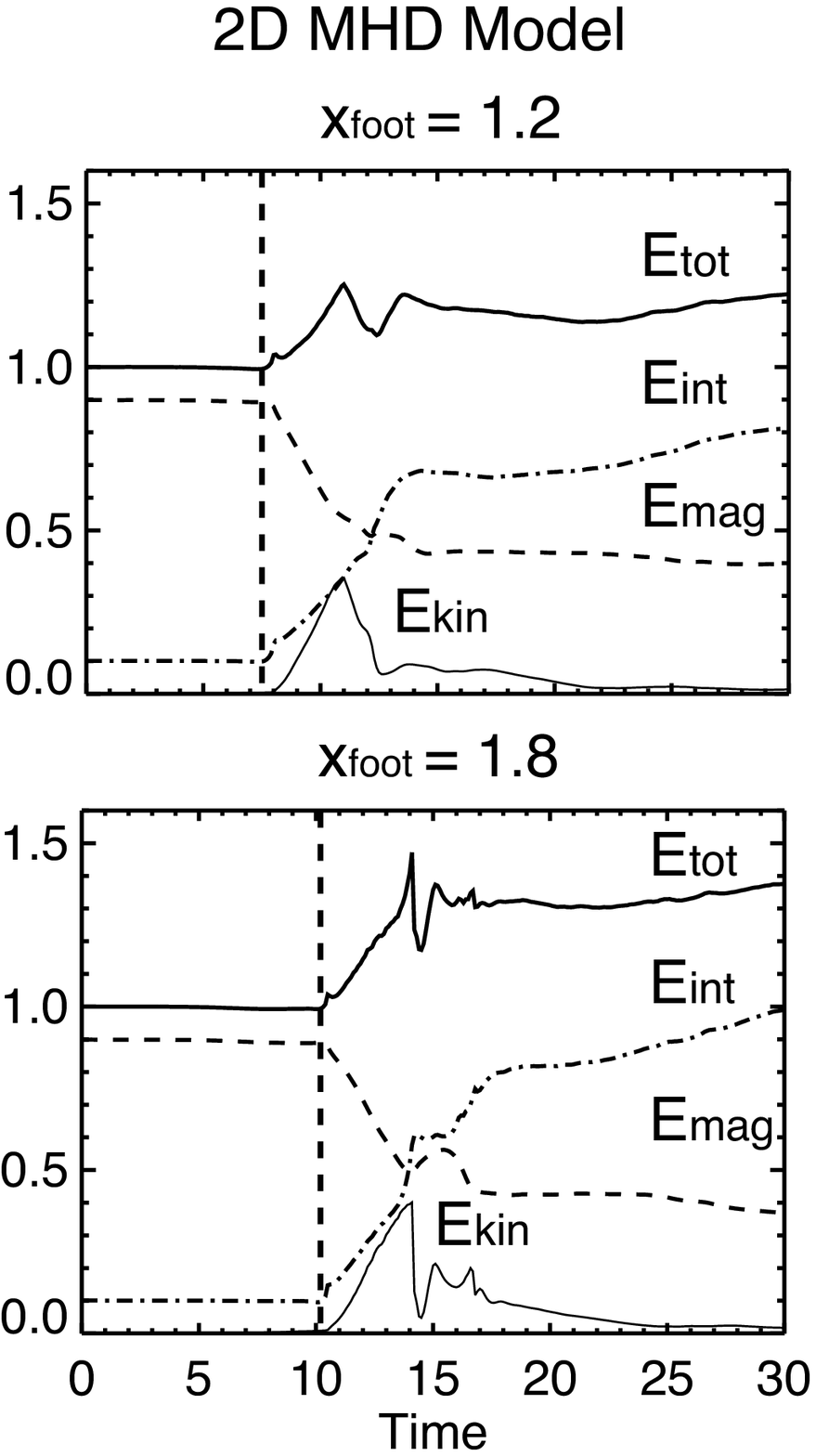}
\caption{Time evolution of the total (thick solid), magnetic (dashed), internal (dash-dot), and kinetic (thin solid) energies along the specified field lines in the 2D MHD model. The vertical dashed lines denote the times when each field line reconnects.\label{fig:energy_line_2dmhd}}
\end{figure}

\begin{figure}
\epsscale{.65}
\plotone{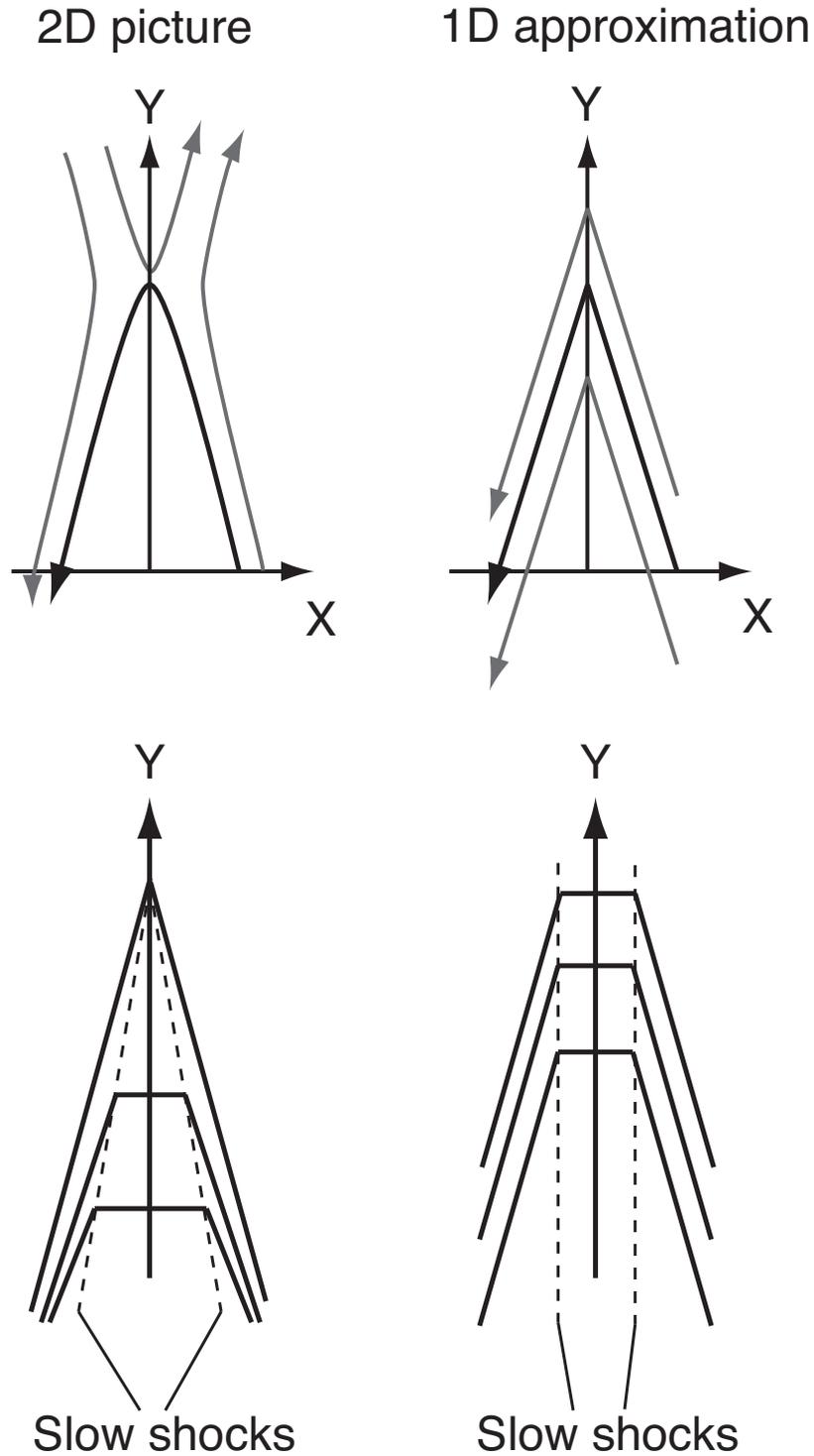}
\caption{Schematic diagram of the reconnected magnetic field in 2D (Left) and the magnetic field mimicking the reconnected field in the 1D-approximation model (pseudo-2D MHD model, Right). The bottom panels show the difference in the reconnected magnetic field configurations between two models. \label{fig:1d_concept}}
\end{figure}

\begin{figure}
\epsscale{.80}
\plotone{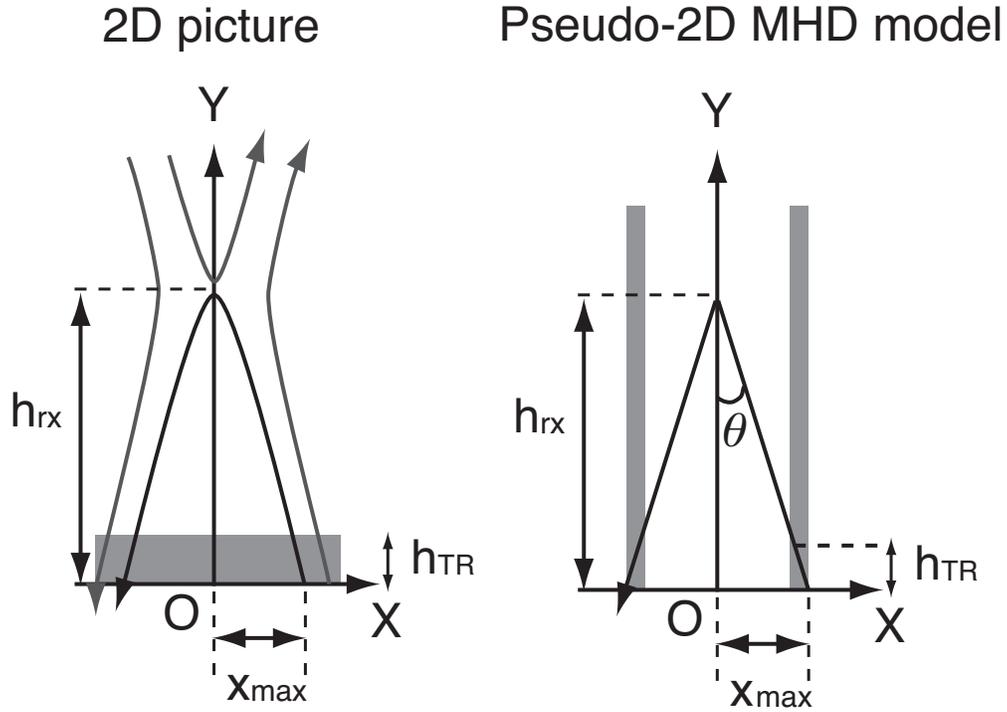}
\caption{Schematic diagram of the set up of our pseudo-2D flare model (Right) based on a 2D picture (Left). $h_{rx}$ and $h_{TR}$ are the height of the reconnection point and height of the transition region, respectively. $\tan^{-1}{(x_{max}/h_{rx})}=\theta$ is the reconnection angle. The grey shaded regions indicate the dense cool material representing the chromosphere. Note that in the pseudo-2D MHD model, all the physical quantities are function of $x$. \label{fig:2d_vs_1d}}
\end{figure}

\begin{figure}
\epsscale{.60}
\plotone{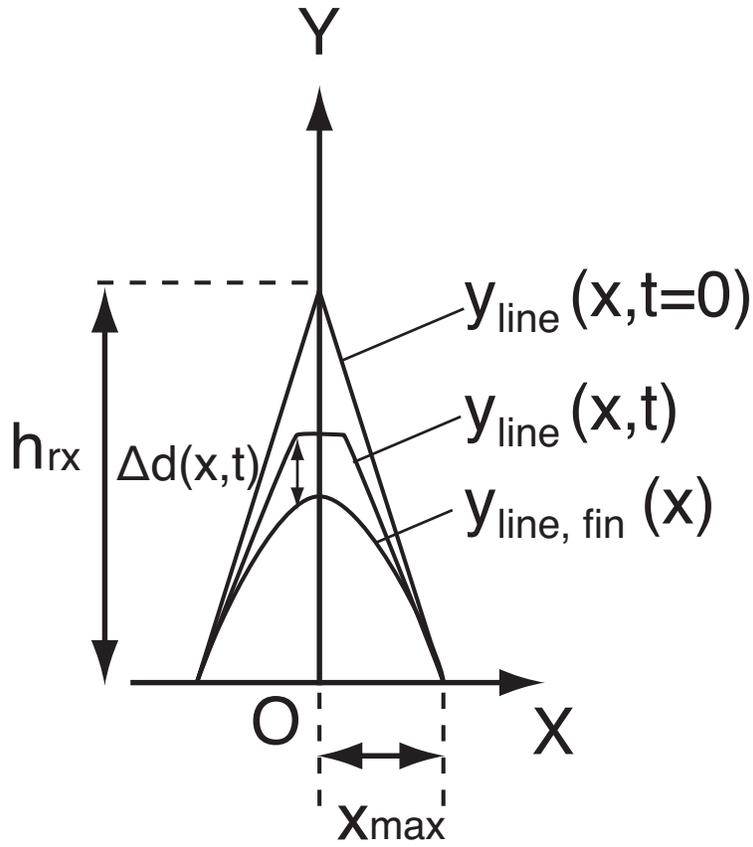}
\caption{The magnetic configurations of a specific field line in the initial state ($y_{line}(x,t=0)$), at $t=t$ ($y_{line}(x,t=t)$), and in the final state ($y_{line,fin}(x)$), respectively. $\Delta d(x,t) = y_{line}(x,t)-y_{line,fin}(x)$ is also described. \label{fig:def_deltad}}
\end{figure}

\begin{figure}
\epsscale{.50}
\plotone{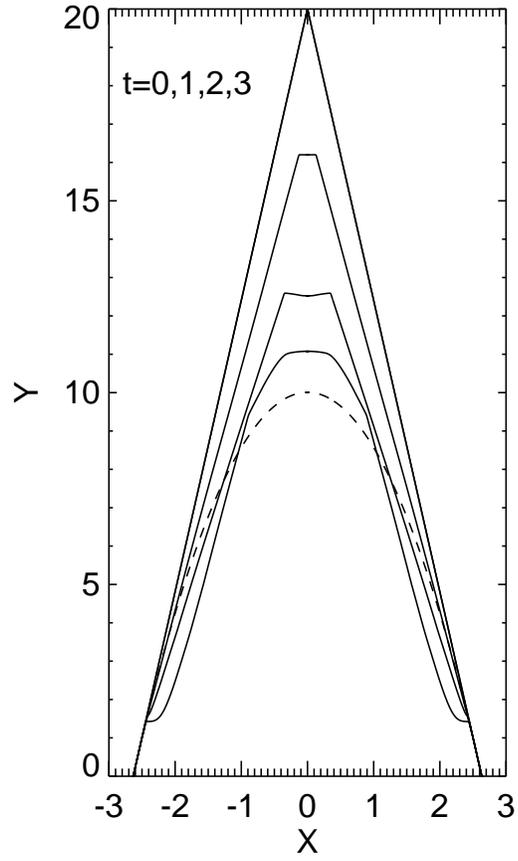}
\caption{Downward moving magnetic field line in the pseudo-2D MHD model. The dashed line denotes the assumed final configuration $y_{line,fin}$. \label{fig:line_evo_p2dmhd}}
\end{figure}

\begin{figure}
\epsscale{.70}
\plotone{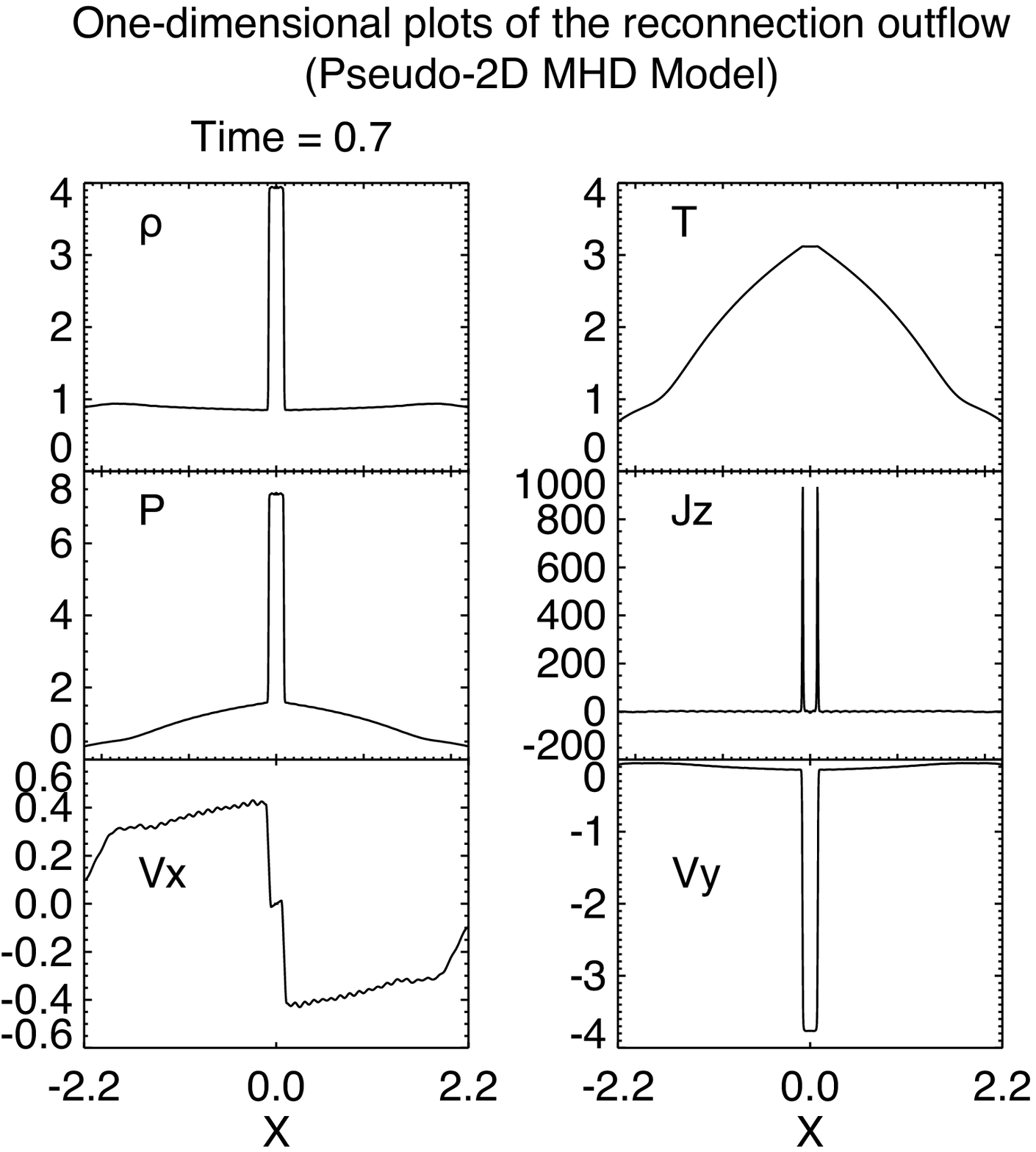}
\caption{One-dimensional plots parallel to the $x$-axis at $t=0.7$ (across the reconnection outflow region). The density, temperature, pressure, $J_z$, $v_x$ and $v_y$ are displayed.  \label{fig:shock_xcut_p2dmhd}}
\end{figure}

\begin{figure}
\epsscale{1.10}
\plotone{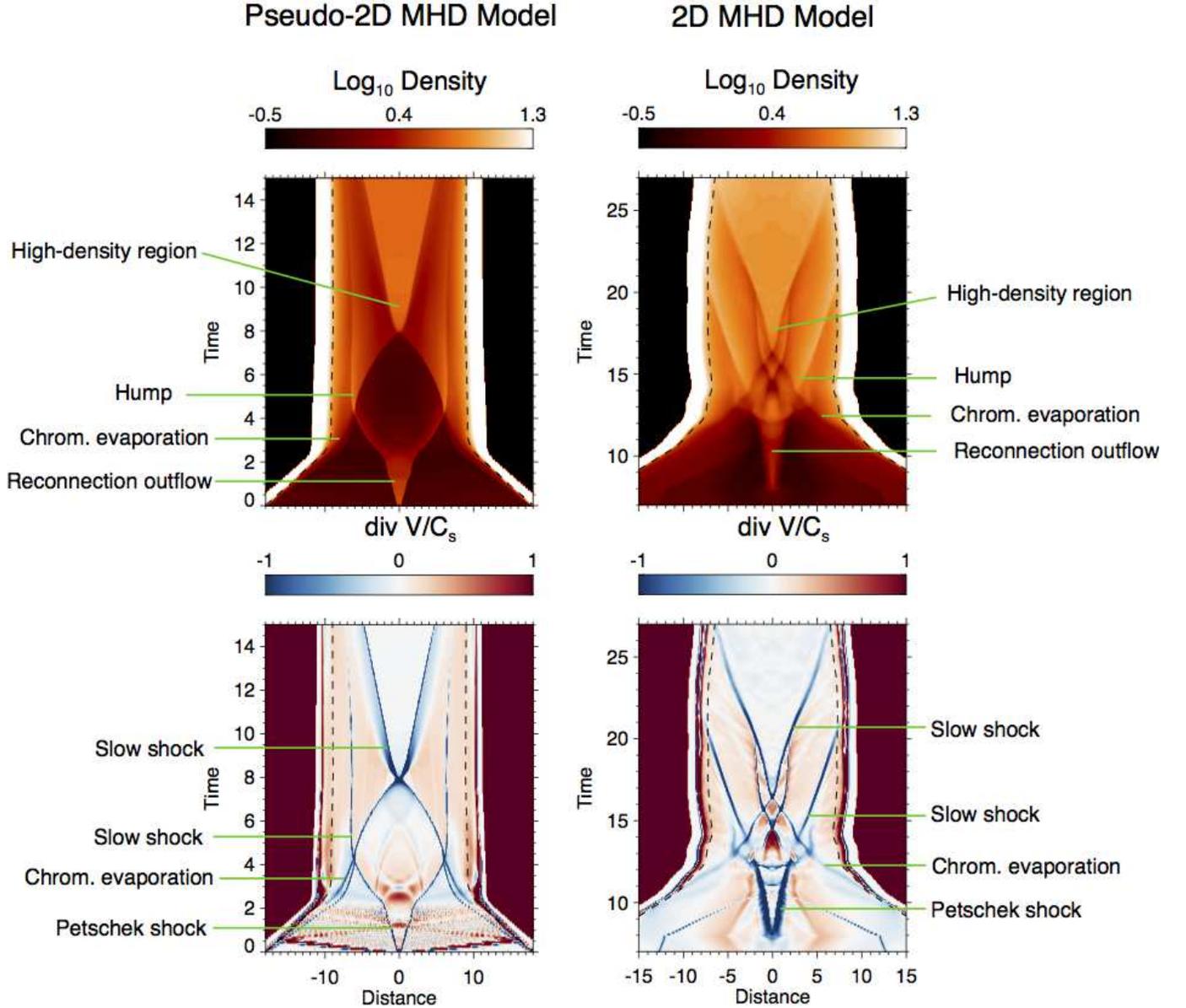}
\caption{Comparison between the pseudo-2D MHD model (Left) and the 2D MHD model (Right). Time-distance diagrams of the density and $\nabla\cdot \bm{v}/C_s$ are shown. The physical quantities of the 2D MHD model are measured along the field line which is originated from $(x,y)=(1.2,0)$. The distance is measured along the field line, and its origin is the apex ($x=0$). The data of the domain of $x<0$ is also shown just for visual inspection. Note that most of the regions where $\nabla \cdot \bm{v}/C_s$ takes large negative values are slow or fast shocks.\label{fig:ro_divv_line_compare}}
\end{figure}

\begin{figure}
\epsscale{.70}
\plotone{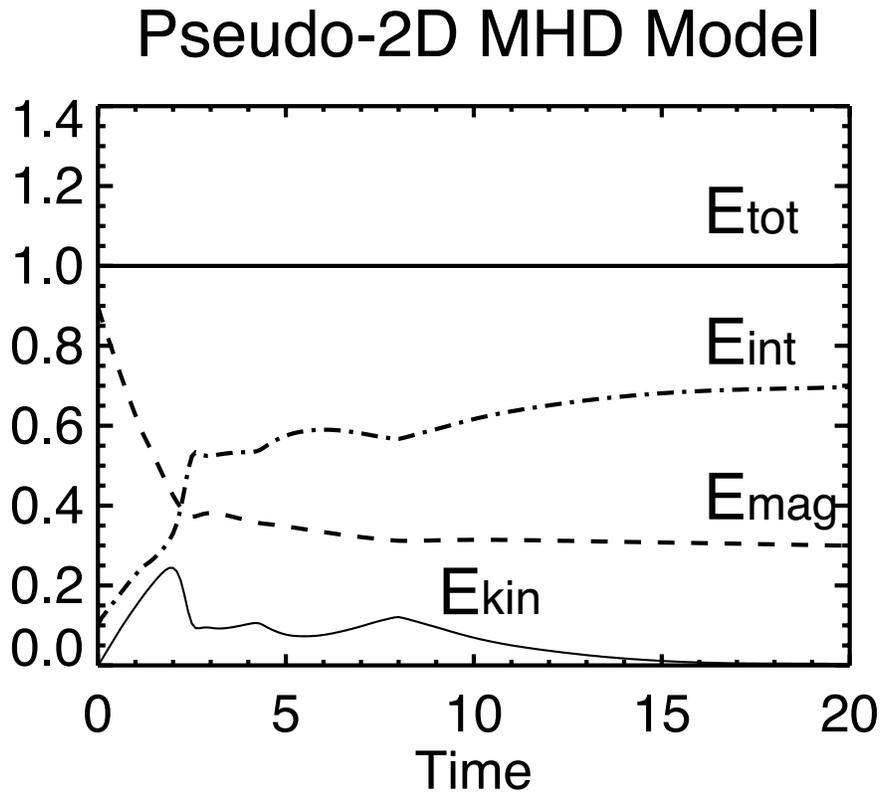}
\caption{Time evolution of the total (thick solid), magnetic (dashed), internal (dash-dot), and kinetic (thin solid) energies along a field line in the pseudo-2D MHD model. \label{fig:energy_line_p2dmhd}}
\end{figure}

\begin{figure}
\epsscale{.30}
\plotone{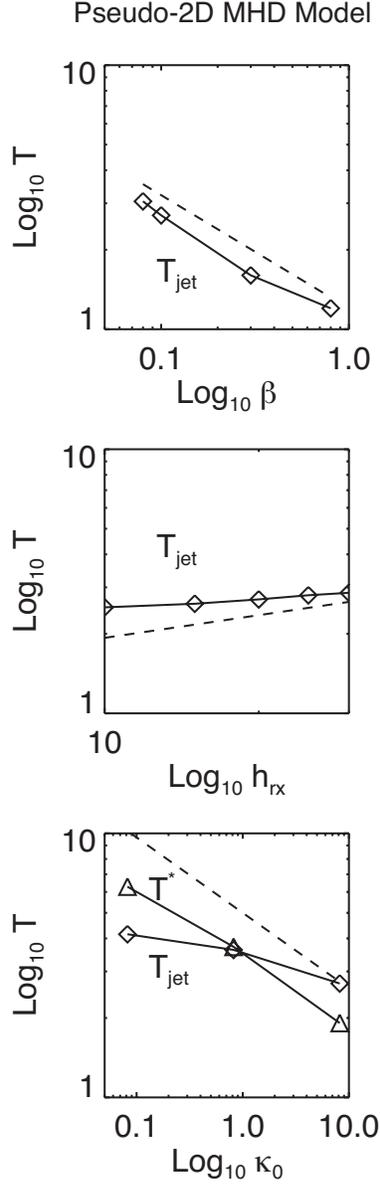}
\caption{Temperature as a function of the initial $\beta$ (Top), the height of the reconnection point $h_{rx}$ (Middle), and the heat conductivity $\kappa_0$ (Bottom), respectively. The data points are taken from the pseudo-2D MHD simulations. $T_{jet}$ denotes the temperature in the reconnection outflow. $T^{*}$ denotes the maximum temperature after the pair of the two slow shocks generated by the chromospheric evaporation flows collides at the apex. The dashed lines show the analytical scaling laws.
 \label{fig:scaling}}
\end{figure}

Performing a 2D MHD simulation, we found new shock structures. The termination shock consists of two oblique fast-mode shocks and sometimes a horizontal fast-mode shock (Figure~\ref{fig:pr_top}). A hump is formed as a result of the collision of the downflow and the chromospheric evaporation flow. After the collision, the fronts of the evaporation flow and downflow become slow shocks, and then the hump appears as a dense region behind the two slow shocks (Figure~\ref{fig:ro_hump_2dmhd}). The upward slow shock finally interact with the slow shock coming from the other side at the top, forming the high-density region. Note that the high-density region is separated from the blob in \citet{yokoyama2001} and is formed below it (Figure~\ref{fig:ro_looptop_2dmhd}). 

We found that the strength of the termination shock in the 2D MHD model shows a quasi-periodic oscillation, which is a multi-dimensional feature. In addition, the shock reflection and Mach reflection are sometimes observed in a concave magnetic structure, which could be important for understanding the heating in the loop-top. These complicated structures at the top will be detailed in our future papers.

We observe no prominent shocks nor waves propagating back and forth from end to end. In addition, no prominent standing slow-mode waves are observed within the calculated time range. By performing 2D MHD simulations without the heat conduction, we confirmed that without the heat conduction the slow shocks formed at the fronts of the downflows from the top propagate back and forth from end to end. Thus the propagation of the slow shocks in the post-flare loops is found to be significantly affected by the heat conduction and evaporation flows. The slow shocks are damped by the heat conduction. The propagation speed of the slow shocks is reduced by the evaporation flow (Doppler effect), which makes the shock propagation time longer. Therefore it is essential to consider the flows resulting from reconnection (particularly downflows from the top) for understanding the behavior of magneto-acoustic waves in the post-flare loops.

In fully 3D situations, reconnection can be intermittent in space and time, which could affect the shock structures found in this study. 3D component reconnection, where reconnecting magnetic field lines are not perfectly anti-parallel, can result in the reconnection outflow jet with a speed insufficient for the formation of the fast shock above the loop top. However, in the case that the reconnection outflow speed exceeds the fast mode phase speed in the outflow region, we expect the formation of the termination shock found as in our 2D MHD model. We also expect that the slow shocks presented in the 2D MHD model will be formed in 3D, since the field-aligned plasma flows which are essential to form slow shocks are well treated in our MHD flare modeling scheme.

From the 2D simulation, we found that the field-aligned plasma motions (particularly evaporation flows and slow shocks) and heat conduction mainly determine the dynamics in the post-flare loops. Considering this, we construct the pseudo-2D MHD model which is basically a 1D MHD model. The pseudo-2D MHD model is compared with the 2D MHD model, and we found that the dynamics (particularly flow structure and shock formation) and the energetics are similar between the two models. The scaling law for the temperature based on a reconnection model is also examined, and it is found that the scaling law holds in the pseudo-2D MHD model. These facts indicate that our pseudo-2D MHD model captures important features of a reconnection model of solar flares.

1D hydrodynamic models, like \citet{mariska1989} and \citet{hori1997,hori1998}, have been used for the post-flare loop modeling. The models are useful to study the thermal evolution and flows in the flare loops, but the energy input (in many cases the heat input) must be done by hand. Our model includes many features of the multi-dimensional MHD processes related to magnetic reconnection, like the heating by the Petschek slow shocks and conversion of the kinetic energy of the reconnection outflow to the heat. Another important point is that our model can treat MHD waves and shocks generated in the post-flare loops, which could be important to understand the flow structure in the loops. 

We note that all the previous 1D hydrodynamic models lack the strong downflows from the top (for example, see Figure~5 in \citet{hori1997}). We showed that the downflows play important roles in forming the humps and slow shocks (Figures~\ref{fig:ro_hump_2dmhd} and \ref{fig:ro_looptop_2dmhd}). Our pseudo-2D MHD model naturally produces the downflows from the top, allowing us to study the dynamic evolution of the thermal structure.

A theoretical model in which field lines shorten after localized 3D reconnection based on the thin flux tube approximation was proposed by \citet{longcope2009}. The model assumes that the plasma has always low-$\beta$. Our 2D MHD simulation demonstrates that, because of the shock heating and compression, the plasma $\beta$ becomes larger than unity not only in the reconnection outflow but also in a large part of the post-flare loops (see Figure~\ref{fig:ro_te_ent_beta_2dmhd}). In such regions, we need to consider the effects of the gas pressure to understand the flow and shock structures. Our pseudo-2D MHD model can treat the high-$\beta$ plasma. \citet{longcope2009} model could be valid in the situation in which the shock heating does not break the low-$\beta$ assumption (e.g. in the situation in which the guide-field is much stronger than the reconnection field). Regarding this issue, see also Appendix~\ref{ap:B}.

The high-$\beta$ condition could lead to disordering of the magnetic field when turbulence is important so that the geometrical assumptions made for our pseudo-2D MHD model are not met. We found no evidence that turbulence is important in the current sheet in our 2D MHD model. Our model will be useful to model the reconnected field lines in such a situation in which turbulence is not important.

As a result of the approximations made for the pseudo-2D MHD model, some multi-dimensional processes such as termination shock formation and turbulence cannot be modeled. However, the pseudo-2D MHD model are able to treat plasma flows and waves/shocks (slow-mode waves/shocks and Alfven waves) along the magnetic field, which could be useful to understand the plasma motions in the post-flare loops.

Our pseudo-2D MHD model requires much smaller computational cost than other multi-dimensional MHD models. This model will allow us to study the evolution of the flare loops in a wide parameter space without expensive computational cost. Also, it will be much easier to include detailed physics like the non-equilibrium ionization effect \citep[e.g.][]{imada2011}. These will be our future work.

\acknowledgments
The authors are grateful to Dr. Seiji Zenitani for fruitful comments on shocks found in our simulations.
ST acknowledges support by the Research Fellowship of the Japan Society for the Promotion of Science (JSPS). This work was supported by a Grant-in-Aid from the Ministry of Education, Culture, Sports, Science and Technology of Japan (No. 25287039).

{\it Facilities:}

\appendix
\section{Numerical Method of Heat Conduction in 2D MHD Model}\label{ap:A}
\begin{figure}
\epsscale{.70}
\plotone{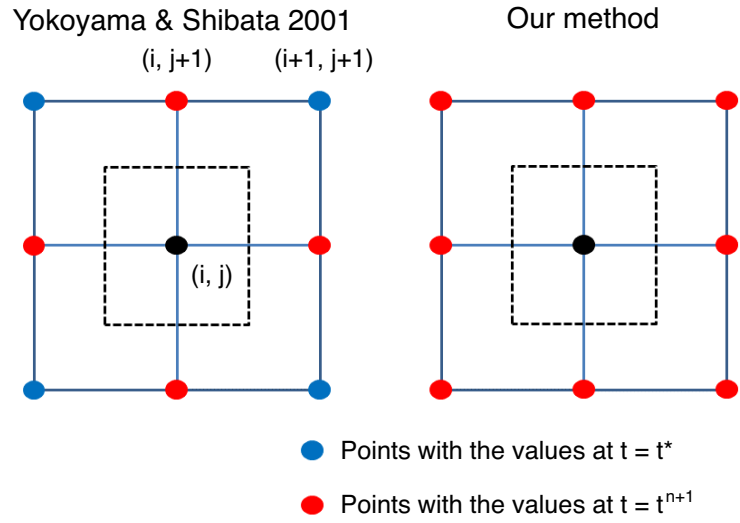}
\caption{The grid points used to calculate the heat flux for the grid $(i,j)$ in \citet{yokoyama2001} (Left) and our method (Right). The red and blue grids have the physical quantities at the time $t=t^{n+1}$ and those at $t=t^*$, respectively. \label{fig:heatcnd}}
\end{figure}

We modify the time step splitting method to calculate the heat conduction flux by \citet{yokoyama2001}.
Using the MHD energy flux $F_{mhd}$ and heat conduction energy flux $F_c$, we can write the energy equation as follows:
\begin{eqnarray}
  \frac{\partial E}{\partial t} + \nabla \cdot \bm{F}_{mhd} + \nabla \cdot \bm{F}_c &=& 0,
\end{eqnarray}
where $E$ is the total energy. 
The discretized form is 
\begin{eqnarray}
  \frac{1}{\Delta t} (E^{n+1}-E^{n})+(\nabla \cdot \bm{F}_{mhd})^{n+1/2}+(\nabla \cdot \bm{F}_c)^{n+1/2} &=& 0.
\end{eqnarray}
First, we calculate the MHD part 
\begin{eqnarray}
  \frac{1}{\Delta t} (E^*-E^n) &=& -(\nabla \cdot \bm{F}_{mhd})^{n+1/2},
\end{eqnarray}
where the superscript $*$ denotes the results of the MHD step. 
Then we calculate the heat conduction step
\begin{eqnarray}
  \frac{1}{\Delta t} (E^{n+1}- E^* ) &=& -(\nabla \cdot \bm{F}_c)^{n+1/2}.
\end{eqnarray}
The procedure mentioned above is the same as \citet{yokoyama2001}. The difference appears in the expression of the heat flux formula. The heat flux formulae in \citet{yokoyama2001} are 
\begin{eqnarray}
 \bm{F}_{c} &\approx&-\kappa_0 (T^*)^{5/2} \frac{\bm{B}^*}{B^*} (\frac{\bm{B}^*}{B^*} \cdot \nabla T), \\
F_{c,x} &\approx& -A_{xx}^* \frac{\partial T}{\partial x}- A_{xy}^* \frac{\partial T}{\partial y}, \\
F_{c,y} &\approx& -A_{yy}^* \frac{\partial T}{\partial y}- A_{yx}^* \frac{\partial T}{\partial x}, \\
A_{ab}^* &=& \kappa_0 (T^*)^{5/2} \frac{B_a^* B_b^*}{(B^*)^2},
\end{eqnarray}
where the subscripts $a$ and $b$ denote $x$ and $y$.
The time and space discretization are, for example,
\begin{eqnarray}
 F_{c,x} &\approx& -A_{xx}^* (\frac{\partial T}{\partial x})^{n+1} -A_{xy}^* (\frac{\partial T}{\partial y})^*.\label{eq:b8}
\end{eqnarray}
We modify this to
\begin{eqnarray}
  F_{c,x} &\approx& -A_{xx}^* (\frac{\partial T}{\partial x})^{n+1} -A_{xy}^* (\frac{\partial T}{\partial y})^{n+1}.\label{eq:b9}
\end{eqnarray}
The difference of two formulae appears in the operator matrix of the implicit scheme used. The number of the non-zero components in one row is 5 in Equation~(\ref{eq:b8}), and 9 in Equation~(\ref{eq:b9}).
Geometrically, in Equation~(\ref{eq:b9}) we use all the neighboring 8 grids around each point at $(n+1)$ step to calculate the heat flux (Figure~\ref{fig:heatcnd}). This method gives more accurate results than the previous method, particularly in the grid points where a magnetic field is largely bend and oblique to coordinate.

\section{Analytical Approach to the Riemann Problem: Possible Solutions}\label{ap:B}
We shall describe the exact solutions of the symmetric MHD Riemann problems that can appear in our pseudo-2D MHD model. In a symmetric MHD Riemann problem, the initial states in the left and right hand side are assumed
to be as follows.
\begin{eqnarray}
 \rho _L &=& \rho _R, \\
 p_L &=& p_R, \\
 B_{x;L} &=&  B_{x;R}, \\
 B_{y;L} &=& -B_{y;R}, \\
 B_{z;L} &=&  B_{z;R}, \\
 \mathbf{v}_R &=& \mathbf{v}_L = 0,
\end{eqnarray}
where the subscripts $L$ and $R$ represent the left hand side and the
right hand side, respectively. 

Let us consider the situation without the guide field
component ($B_{z;L} =  B_{z;R}=0$).
Since the magnetic field vectors in the initial states are in the $x$-$y$ plane, the rotational discontinuities will not appear in the solutions.
Due to the initial discontinuity, one contact discontinuity, two slow shocks, and two fast
rarefaction waves are generated.

\begin{figure}
\epsscale{.7}
\plotone{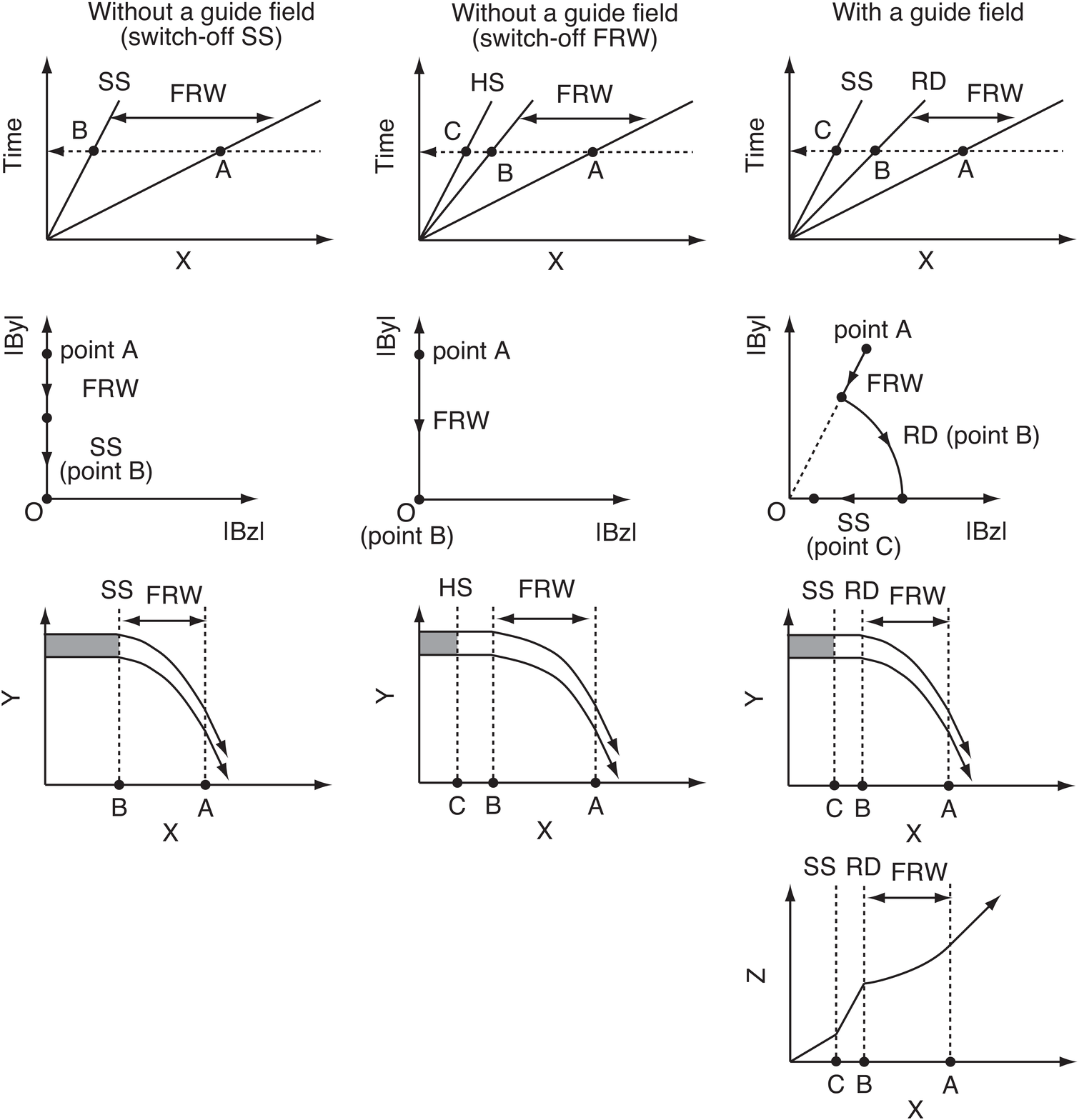}
\caption{Three possible solutions of the Riemann problem. Left column: Case without a guide field, where $B_y$ is switched off by the slow shock (SS). Middle column: Case without a guide field, where $B_y$ is switched off by the fast mode rarefaction wave (FRW). Right column: Case with a guide field, where the magnetic field is rotated by the rotational discontinuity (RD) and the magnetic energy is released at the slow shock (SS). First row: $x-t$ diagrams that show the Riemann fans. Second row: $|B_z|-|B_y|$ diagrams that show the transitions through the regions in the Riemann fans (the paths are indicated by the dashed lines in the first row). The third and fourth rows: $x-y$ diagrams that show the magnetic field configurations on the plane. Only for the case with a guide field, $x-z$ diagram is also depicted. The grey regions indicate the high-entropy regions due to the shock dissipation.  \label{fig:wave_structure}}
\end{figure}

From the jump condition across the contact
discontinuity, the tangential magnetic field, $B_y$, should be equal to zero in
the region between two slow shocks. Therefore, it is necessary to have
either  the switch-off slow shock (SS) or the switch-off fast rarefaction wave (FRW).
In the former case (see the left panels of Figure~\ref{fig:wave_structure}), the exact solution consists of the switch-off SS and FRW. On the other hand, in the latter case (the middle panels of Figure \ref{fig:wave_structure}), the exact solution consists of the pure hydro shock (HS) and switch-off FRW.

\begin{figure}
\epsscale{.80}
\plotone{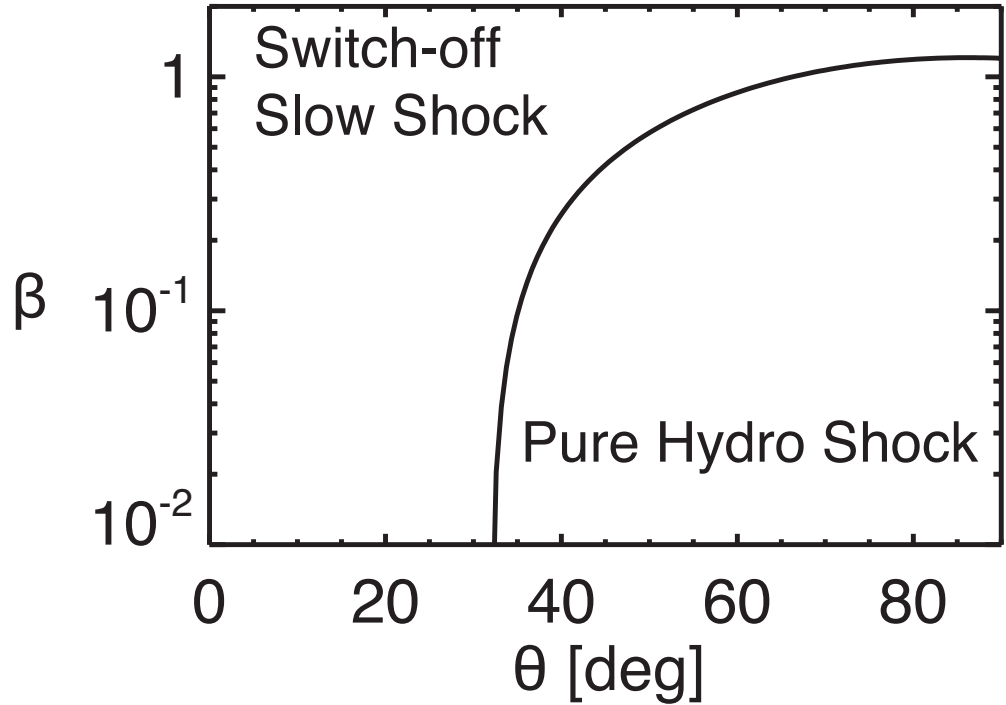}
\caption{A phase diagram of the possible solutions of the symmetric Riemann problem in the situation without a guide-field. $\theta$ is the reconnection angle and $\beta$ is the plasma beta. "Switch-off slow shock" regime and "pure hydro shock" regime corresponds to the left and middle panels in Figure~\ref{fig:wave_structure}, respectively.\label{fig:divide_region}}
\end{figure}

There are two parameters in the symmetric MHD Riemann problems: plasma
beta $\beta$ and the reconnection angle $\theta$.  A sophisticated procedure to derive the exact solutions of more general MHD Riemann problems was developed by \citet{falle1998}. We utilize the procedure to derive the exact solutions of our problems. The result is summarized in the phase diagram of Figure~\ref{fig:divide_region}, where $\beta$ and $\theta$ are the key parameters. Here the heat conduction is neglected. The $\theta$-$\beta$ space is divided into two regions: switch-off slow shock regime and pure hydro shock regime. The parameter set of the example introduced in this paper is within the switch-off slow shock region. If we include the heat conduction effect, the pure hydro shock region will become slightly wider.

\begin{figure}
\epsscale{.80}
\plotone{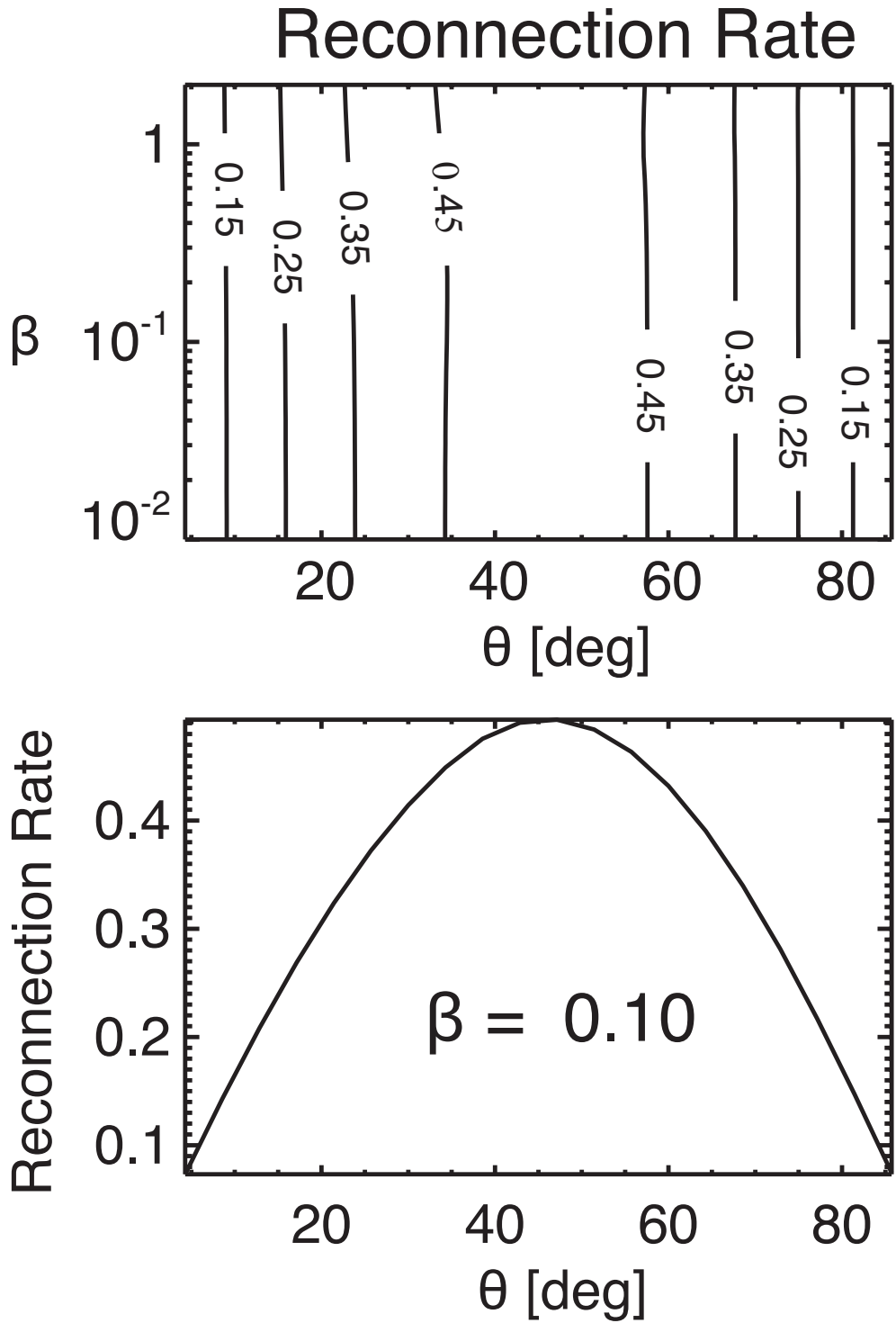}
\caption{Top: The normalized reconnection rate in the $\theta$-$\beta$ plane. Bottom: The normalized reconnection rate for $\beta=0.1$ as a function of $\theta$. \label{fig:reconrate_1d}}
\end{figure}

Figure~\ref{fig:reconrate_1d} displays the normalized reconnection rate in the $\theta$-$\beta$ diagram, where the normalized reconnection rate is defined by $v_{y}B_x/(V_{A}B_{x0})$. Where $v_{y}$ is the outflow speed, $V_A=\sqrt{B_x^2+B_y^2}/\sqrt{\rho}$, and $B_x=B_{x0}$. The maximum reconnection rate observed in solar flares and simulations is of the order of 0.1. To take the reconnection rate similar to 0.1 in our model, we need to choose a reasonable parameter set from this diagram.

If a guide field exists, the rotational discontinuities (RD) will appear in addition to SS and FRW.
In this case it is necessary to have either the switch-off SS, switch-off FRW, or RD because of the boundary condition at the contact discontinuity ($B_y=0$). Among them, only the solution with RD (the right panels of
Figure~\ref{fig:wave_structure}) meets the requirement that the continuity condition of $V_z$ at the contact discontinuity should be satisfied \citep{petschek1967}. \citet{longcope2014} claimed that pure hydro shocks can be formed in a reconnected flux tube even when a guide field exists. However, considering the analysis here, pure hydro shocks will not appear when a guide field exists.

\clearpage

\begin{table}
\begin{center}
\caption{Normalization units\label{tab:units}}
\begin{tabular}{crr}
\tableline\tableline
Quantity & Unit & Value\\
\tableline
Length & $L_0$ & 3,000~km\\
Velocity & $C_{s0}=\left[ \gamma(k_B/m)T_{0} \right]^{1/2}$ & 170~km~s$^{-1}$ \\
Time & $L_0/C_{s0} $ & 18~s \\
Temperature & $T_0=T_{cr}$ & $2\times 10^6$~K\\
Density & $\rho_0=\rho_{cr}$ & (10$^9$~cm~$^{-3}$)$\times m$\\
Pressure & $\gamma (k_B/m)\rho_0 T_0$ & 0.47~dyn~cm$^{-2}$\\
Magnetic field strength & $\left[8\pi \gamma (k_B/m)\rho_0 T_0 \right]^{1/2}$ & 3.4~G\\
\tableline
\end{tabular}
\end{center}
\end{table}

\clearpage

\clearpage

\clearpage

\clearpage

\clearpage

\clearpage

\clearpage

\clearpage

\begin{thebibliography}{}
\bibitem[Chen et al.(1999)]{chen1999} Chen, P.~F., Fang, C., Tang, Y.~H., \& Ding, M.~D.\ 1999, \apj, 513, 516 
\bibitem[Drake et al.(2006)]{drake2006} Drake, J.~F., Swisdak, M., Che, H., \& Shay, M.~A.\ 2006, \nat, 443, 553
\bibitem[Falle et al.(1998)]{falle1998} Falle, S.~A.~E.~G., Komissarov, S.~S., \& Joarder, P.\ 1998, \mnras, 297, 265 
\bibitem[Fisher et al.(1985)]{fisher1985} Fisher, G.~H., Canfield, R.~C., \& McClymont, A.~N.\ 1985, \apj, 289, 414 
\bibitem[Fisher \& Hawley(1990)]{fisher1990} Fisher, G.~H., \& Hawley, S.~L.\ 1990, \apj, 357, 243 
\bibitem[Forbes et al.(1989)]{forbes1989} Forbes, T.~G., Malherbe, J.~M., \& Priest, E.~R.\ 1989, \solphys, 120, 285 
\bibitem[G{\"u}del(2004)]{gudel2004} G{\"u}del, M.\ 2004, \aapr, 12, 71
\bibitem[Hayashi et al.(1996)]{hayashi1996} Hayashi, M.~R., Shibata, K., \& Matsumoto, R.\ 1996, \apjl, 468, L37
\bibitem[Hori et al.(1997)]{hori1997} Hori, K., Yokoyama, T., Kosugi, T., \& Shibata, K.\ 1997, \apj, 489, 426 
\bibitem[Hori et al.(1998)]{hori1998} Hori, K., Yokoyama, T., Kosugi, T., \& Shibata, K.\ 1998, \apj, 500, 492 
\bibitem[Imada et al.(2011)]{imada2011} Imada, S., Murakami, I., Watanabe, T., Hara, H., \& Shimizu, T.\ 2011, \apj, 742, 70 
\bibitem[Kigure et al.(2010)]{kigure2010} Kigure, H., Takahashi, K., Shibata, K., Yokoyama, T., \& Nozawa, S.\ 2010, \pasj, 62, 993 
\bibitem[Koyama et al.(1994)]{koyama1994} Koyama, K., Maeda, Y., Ozaki, M., et al.\ 1994, \pasj, 46, L125 
\bibitem[Krucker et al.(2010)]{krucker2010} Krucker, S., Hudson, H.~S., Glesener, L., et al.\ 2010, \apj, 714, 1108 
\bibitem[Longcope et al.(2009)]{longcope2009} Longcope, D.~W., Guidoni, S.~E., \& Linton, M.~G.\ 2009, \apjl, 690, L18
\bibitem[Longcope(2014)]{longcope2014} Longcope, D.~W.\ 2014, \apj, 795, 10 
\bibitem[Machida \& Matsumoto(2003)]{machida2003} Machida, M., \& Matsumoto, R.\ 2003, \apj, 585, 429 
\bibitem[Mariska et al.(1989)]{mariska1989} Mariska, J.~T., Emslie, A.~G., \& Li, P.\ 1989, \apj, 341, 1067 
\bibitem[Masada et al.(2010)]{masada2010} Masada, Y., Nagataki, S., Shibata, K., \& Terasawa, T.\ 2010, \pasj, 62, 1093 
\bibitem[Masuda et al.(1995)]{masuda1995} Masuda, S., Kosugi, T., Hara, H., et al.\ 1995, \pasj, 47, 677 
\bibitem[McKenzie(2013)]{mckenzie2013} McKenzie, D.~E.\ 2013, \apj, 766, 39 
\bibitem[Miyoshi \& Kusano(2005)]{miyoshi2005} Miyoshi, T., \& Kusano, K.\ 2005, Journal of Computational Physics, 208, 315 
\bibitem[Miyagoshi \& Yokoyama(2004)]{miyagoshi2004} Miyagoshi, T., \& Yokoyama, T.\ 2004, \apj, 614, 1042 
\bibitem[Nagai(1980)]{nagai1980} Nagai, F.\ 1980, \solphys, 68, 351 
\bibitem[Nakariakov et al.(2004)]{nakariakov2004} Nakariakov, V.~M., Tsiklauri, D., Kelly, A., Arber, T.~D., \& Aschwanden, M.~J.\ 2004, \aap, 414, L25 
\bibitem[Nishizuka \& Shibata(2013)]{nishizuka2013} Nishizuka, N., \& Shibata, K.\ 2013, Physical Review Letters, 110, 051101 
\bibitem[Ofman \& Wang(2002)]{ofman2002} Ofman, L., \& Wang, T.\ 2002, \apjl, 580, L85 
\bibitem[Oka et al.(2015)]{oka2015} Oka, M., Krucker, S., Hudson, H.~S., \& Saint-Hilaire, P.\ 2015, \apj, 799, 129 
\bibitem[Peres \& Reale(1993)]{peres1993} Peres, G., \& Reale, F.\ 1993, \aap, 275, L13 
\bibitem[Petschek \& Thorne(1967)]{petschek1967} Petschek, H.~E., \& Thorne, R.~M.\ 1967, \apj, 147, 1157 
\bibitem[Seaton \& Forbes(2009)]{seaton2009} Seaton, D.~B., \& Forbes, T.~G.\ 2009, \apj, 701, 348 
\bibitem[Shibata \& Yokoyama(2002)]{shibata2002} Shibata, K., \& Yokoyama, T.\ 2002, \apj, 577, 422
\bibitem[Shibata \& Magara(2011)]{shibata2011} Shibata, K., \& Magara, T.\ 2011, Living Reviews in Solar Physics, 8, 6 
\bibitem[Shiota et al.(2005)]{shiota2005} Shiota, D., Isobe, H., Chen, P.~F., et al.\ 2005, \apj, 634, 663 
\bibitem[Somov \& Kosugi(1997)]{somov1997} Somov, B.~V., \& Kosugi, T.\ 1997, \apj, 485, 859 
\bibitem[Tanuma \& Shibata(2005)]{tanuma2005} Tanuma, S., \& Shibata, K.\ 2005, \apjl, 628, L77 
\bibitem[Tsuboi et al.(1998)]{tsuboi1998} Tsuboi, Y., Koyama, K., Murakami, H., et al.\ 1998, \apj, 503, 894 
\bibitem[Tsuneta \& Naito(1998)]{tsuneta1998} Tsuneta, S., \& Naito, T.\ 1998, \apjl, 495, L67 
\bibitem[Ugai(1992)]{ugai1992} Ugai, M.\ 1992, Physics of Fluids B, 4, 2953 
\bibitem[V{\"o}gler et al.(2005)]{vogler2005} V{\"o}gler, A., Shelyag, S., Sch{\"u}ssler, M., et al.\ 2005, \aap, 429, 335 
\bibitem[Warren(2006)]{warren2006} Warren, H.~P.\ 2006, \apj, 637, 522 
\bibitem[Yokoyama \& Shibata(1997)]{yokoyama1997} Yokoyama, T., \& Shibata, K.\ 1997, \apjl, 474, L61
\bibitem[Yokoyama \& Shibata(1998)]{yokoyama1998} Yokoyama, T., \& Shibata, K.\ 1998, \apjl, 494, L113 
\bibitem[Yokoyama \& Shibata(2001)]{yokoyama2001} Yokoyama, T., \& Shibata, K.\ 2001, \apj, 549, 1160
\bibitem[Zenitani \& Miyoshi(2011)]{zenitani2011} Zenitani, S., \& Miyoshi, T.\ 2011, Physics of Plasmas, 18, 022105 
\end{thebibliography}
\end{document}